\RequirePackage{lineno}
\documentclass[prb,twocolumn,showpacs,floatfix]{revtex4}
\usepackage[final]{graphicx}
\usepackage{amssymb}
\usepackage{amsmath}
\usepackage{xspace}
\usepackage{url}
\usepackage{wasysym}
\usepackage{color}

\usepackage{xcolor,soul}
\sethlcolor{yellow}

\DeclareMathOperator{\sgn}{sgn}
\newcommand{\phdag}{{\phantom{\dag}}}

\newcommand{\HeliumGroundState}{\mathrm{He^0(1s^2,1^1S_0)}}
\newcommand{\HeliumPositiveIon}{\mathrm{He^+(1s,1^2S_{1/2})}}
\newcommand{\HeliumNegativeIon}{\mathrm{He^{\ast-}(1s2s^2,2^2S_{1/2})}}
\newcommand{\HeliumMetaSing}{\mathrm{He^\ast(1s2s,2^1S_0)}}
\newcommand{\HeliumMetaTrip}{\mathrm{He^\ast(1s2s,2^3S_1)}}

\begin{document} %%%%%%%%%%%%%%%%%%%%%%%%%%%%%%%%%%%%%%%%%%%%%%%%%%%%%%%%%%%%%%
%------------------------------------------------------------------------------
% Title
%------------------------------------------------------------------------------
%\title{Electron emission in charge-transferring collisions of ${\rm He}^+$ ions with metal surfaces}
%\title{Quantum kinetic theory of ion-induced secondary electron emission from surfaces}
\title{Ion-induced secondary electron emission from metal surfaces}
%------------------------------------------------------------------------------
% Authors
%------------------------------------------------------------------------------
%------------------------------------------------------------------------------
% Date
%------------------------------------------------------------------------------
\author{M. Pamperin, F. X. Bronold, and H. Fehske}
\affiliation{Institut f{\"ur} Physik,
             Ernst-Moritz-Arndt-Universit{\"a}t Greifswald,
             17489 Greifswald,
             Germany}
\date{\today}
\begin{abstract}
Using a helium ion hitting various metal surfaces as a model system, we 
describe a general quantum-kinetic approach for calculating ion-induced 
secondary electron emission spectra at impact energies where the emission 
is driven by the internal potential energy of the ion. It is based on an 
effective model of the Anderson-Newns-type for the subset of electronic 
states of the ion-surface system most strongly affected by the collision. 
Central to our approach is a pseudo-particle representation for the electronic
configurations of the projectile which enables us, by combining it with 
two additional auxiliary bosons, to describe in a single Hamiltonian emission 
channels involving electronic configurations with different internal 
potential energies. It is thus possible to treat Auger neutralization of 
the ion on an equal footing with Auger de-excitation of temporarily formed 
radicals and/or negative ions. From the Dyson equations for the projectile 
propagators and an approximate evaluation of the selfenergies rate equations 
are obtained for the probabilities with which the projectile configurations
occur and an electron is emitted in the course of the collision. Encouraging
numerical results, especially for the helium-tungsten system, indicate the 
potential of our approach.
\end{abstract}


\maketitle

\section{Introduction}
\label{Introduction}
In low-temperature gas discharges secondary electron emission from the walls confining
the plasma is an important surface collision process caused by atomic
constituents of the plasma hitting the wall~\cite{LL05}. Known since the early days of 
gaseous electronics~\cite{LM24}, it moved into the focus of interest again quite recently. 
For instance, it has been shown that the ionization dynamics~\cite{SDS11,GNO13}, the 
electron power absorption~\cite{DDW17}, and a number of other quantities and 
processes~\cite{HG16} in capacitively coupled discharges depend significantly on the 
secondary electron emission coefficient, that is, the probability with which an electron 
is released in the course of an atom-surface collision. It has been also demonstrated 
that the structure of the plasma sheath is strongly affected by secondary electron 
emission~\cite{CU16,LW15,SKR09,TLC04}. The impact energies are typically 
in the range where electron emission is driven by the internal 
potential energy stored in the electronic configuration of the projectile. Auger 
neutralization of ions and/or Auger de-excitation of metastable species are 
thus the main channels of secondary electron emission~\cite{PP99}. Depending on the
initial state of the projectile ion- and radical-induced secondary electron emission
can thus be distinguished. Since the processes are also of interest for themselves as well as 
of importance for various kinds of surface diagnostics, for instance, secondary ion mass 
spectroscopy~\cite{CH91} or metastable atom de-excitation spectroscopy~\cite{HMO97}, Auger 
and related charge-transfer processes have been reviewed several 
times~\cite{Monreal14,Winter07,Winter02,Baragiola94,LG90,BN89,Modinos87,YM86,NMB83} 
since the early studies~\cite{OM30,Massey30,Shekhter37,CL44,Hagstrum53,Hagstrum54a} 
dating back to the very beginning of modern condensed matter physics. There can be thus no doubt
that the basic mechanisms of secondary electron emission from surfaces have by now been
identified. 

Although the principles of secondary electron emission are known it is still
a great challenge to measure or to calculate secondary electron emission spectra, even
for free-standing surfaces not in contact with a plasma. Experimentally it requires sophisticated 
instrumentation~\cite{LKW07,LKW03,SHL03,Winter02,HWB98,Winter93,KC93,MHB93,BMN92,SWK87,Propst63}, 
whereas theoretically the challenge is to find an efficient way to deal 
with a many-body scattering problem giving rise to a great variety of collision
pathways~\cite{BGG16,IGG14,IGG13,MGP13,MSS09,VGB05,GWM03,WGM01,SZN00,MMM98,LCG98,CLD98,LM96,LMA94,MAB93,ME87}. 
It is thus not surprising that the data base for secondary electron emission is rather sparse,
especially for materials used as walls in laboratory gas discharges. There have been 
only a few experimental efforts devoted to measure secondary electron emission coefficients 
specifically for them~\cite{DBS16,MCA15}. 

\begin{figure}[t]
        \includegraphics[width=0.95\linewidth]{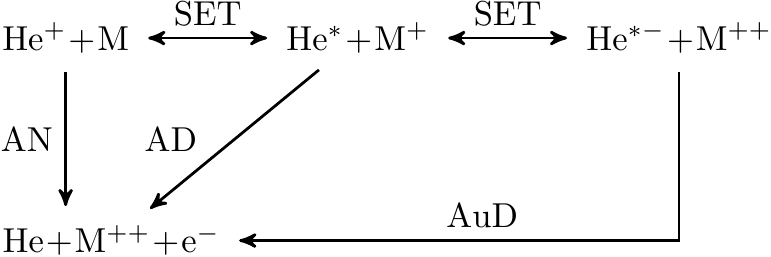}
	\caption{Schematic representation of possible charge transfer processes
		which may take place during a collision of a ${\rm He}^+$ ion with a metal 
                surface, depending on the occupancies of the electronic states, their coupling, 
                and the collision dynamics. The ion may capture electrons from the metal M by 
                single-electron transfer (SET) changing its configuration from $\mathrm{He}^+{\rm (1s)}$ 
		to $\mathrm{He}^\ast {\rm (1s2s)}$ or even to $\mathrm{He}^{\ast-}{\rm (1s2s^2)}$ if two 
                sequential SET processes occur. SET processes may however also work in the other direction, 
                that is, the projectile may also loose electrons. Autodetachment (AuD) may lead to an 
		electron loss and a reconfiguration of $\mathrm{He}^{\ast-}{\rm (1s2s^2)}$ to 
                $\mathrm{He}{\rm (1s^2)}$. In addition, Auger neutralization (AN) and Auger de-excitation 
		(AD) due to the Coulomb interaction between two electrons may take place pushing 
                the projectile to its groundstate configuration $\mathrm{He}{\rm (1s^2)}$ thereby
                also releasing an electron. The charge state of the metal M is indicated 
                to emphasize the charge-transfer taking place due to the various processes.} 
	\label{outlineProcesses}
\end{figure}

To illustrate the complexity of the physics involved 
we show in Fig.~\ref{outlineProcesses} the collision channels which may be open when a positive 
helium ion hits a metal surface and releases an electron. Besides Auger neutralization  
of the positive ion itself, there is a sequence of single-electron transfers possible,
leading to neutral and negatively charged metastable states which may Auger de-excite 
or autodetach to the helium groundstate thereby also releasing an electron. 
Which one of the three channels dominates depends on the collision parameters and the 
metal. An unbiased description of the collision requires thus a theoretical model capable 
to treat all channels having a chance to be involved in the electron emission simultaneously.
To present such a theory is the purpose of this work. 

We do not attempt a description from first principles~\cite{IGG14,IGG13,MGP13,VGB05,GWM03,MMM98}. 
Instead, we use an Anderson-News-Hamiltonian for the subset of electronic degrees of freedom 
which are dominantly involved in the collision process. Combined with Gadzuk's semiempirical 
approach~\cite{Gadzuk67a,Gadzuk67b} of determining the matrix elements of this Hamiltonian 
from classical image shifts it yields a rather flexible basis for the modeling of a great 
variety of projectile-target combinations. We consider this type of effective modeling, 
requiring at the end only a few parameters with a clear physical interpretation, particularly 
appropriate for describing secondary electron emission from plasma walls which are often 
microscopically not well characterized preventing thereby a more sophisticated modeling. 
Local correlations on the projectile can be taken into account by a projection operator 
and pseudo-particle technique pioneered by Langreth and coworkers~\cite{LN91,SLN94a,SLN94b}. 
Combining this with additional auxiliary bosons to accommodate the energy defects 
between different electronic configurations of the projectile leads to a Hamiltonian 
containing as many projectile configurations as one wishes to include and at the same time 
is amenable to a quantum kinetic analysis~\cite{LN91,SLN94a,SLN94b}. At the end, it leads 
to rate equations for the probabilities with which the electronic configurations of the 
projectile occur and an electron is emitted in the course of the collision. We employed 
this approach previously to describe electron emission from metal and dielectric surfaces 
due to the de-excitation of metastable nitrogen molecules~\cite{MBF11,MBF12a,MBF12b} and to the 
neutralization of positive strontium and magnesium ions at gold surfaces~\cite{PBF15a,PBF15b}. 
In this work we apply it to a positive helium ion hitting various metal surfaces. 
Confronted with experimental data~\cite{MHB93,LKW03} the approach turns out to yield secondary 
electron emission coefficients of the correct order of magnitude and may even be able to 
produce the correct shape of the emission spectrum if it was augmented by scattering 
processes~\cite{LKW07,BMN92,Propst63} which we so far however did not include in the model.   

The outline of the remainder of the paper is as follows. In the next section we set up the
Anderson-Newns model for the emission channels shown in Fig.~\ref{outlineProcesses}. Besides 
explaining how the matrix elements of the Hamiltonian are obtained from Gadzuk's reasonings 
we also give the details of the projection operator and pseudo-particle technique which 
enables us to encode into a single Hamiltonian electronic configurations with different 
internal potential energies. Section~\ref{QuantumKinetics} together with an appendix describes
the quantum-kinetic derivation of the rate equations for the probabilities with which the 
various electronic configurations of the projectile are realized in the collision and an 
electron is emitted. Numerical results are presented in Sect.~\ref{Results} and concluding 
remarks summarize and assess our approach in Sect.~\ref{Conclusions}.

\section{Model}
\label{Model}

\begin{figure}[t]
        \includegraphics[width=0.9\linewidth]{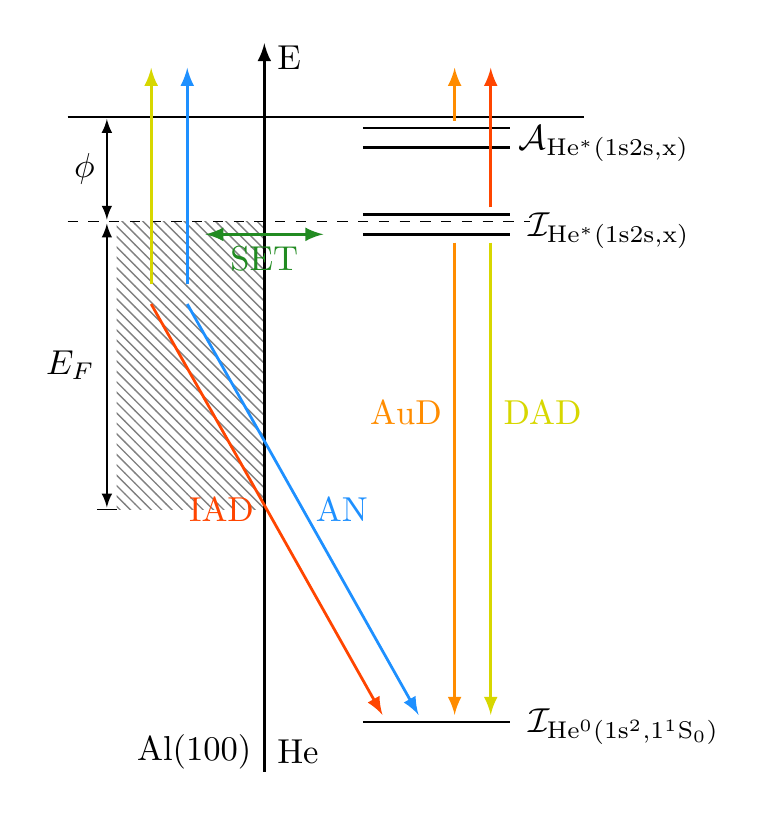}
        \caption{(Color online) On scale representation of the electronic states 
        involved in the neutralization of a $\mathrm{He}^+(1s)$ ion on an  
        $\mathrm{Al}(100)$ surface due to the processes introduced and discussed in 
        Fig.~\ref{outlineProcesses}. The situation shown corresponds to the case 
        where projectile and target are infinitely far apart. Polarization-induced 
        shifts of the ionization energies, $\mathcal{I}_\HeliumGroundState$, 
        $\mathcal{I}_\HeliumMetaSing$, and $\mathcal{I}_\HeliumMetaTrip$ and the 
        electron affinities $\mathcal{A}_\HeliumMetaSing$ and $\mathcal{A}_\HeliumMetaTrip$ 
        encoded in Eqs.~\eqref{1sShift}--\eqref{2sTNshift} are not shown. The
        shaded region on the left indicates the occupied states of the conduction band 
        of the $\mathrm{Al}(100)$ surface, the label "x" in the ionization and affinity 
        levels of the metastable configuration stands for either the triplet or the singlet
        term symbol, and the color-coded arrows give the  
        transitions involved in secondary electron emission due to Auger neutralization
        (AN, blue), direct and indirect Auger de-excitation (DAD, yellow; IAD, red), 
        and autodetachment (AuD, orange), where the latter two take place only after  
        single-electron transfers (SET, green) have occurred.}
        \label{physicalProcesses}
\end{figure}

When an atomic projectile approaches a surface direct and exchange Coulomb 
interactions take place between their individual constituents leading to a 
modification of the projectile's and target's electronic structure. In some 
cases this may cause a redistribution of electrons between them accompanied 
perhaps by an emission of an electron. Since the projectile 
and the target are composite systems, to analyze these processes theoretically is a 
complicated many-body problem. It can be either approached with first-principle 
methods~\cite{IGG14,IGG13,MGP13,VGB05,GWM03,MMM98}, ideally containing the full 
electronic structure of the target and the projectile, or with model 
Hamiltonians focusing only on the subset of electronic states which are actively 
involved in the collision as pioneered by Gadzuk~\cite{Gadzuk67a,Gadzuk67b}. The former 
is computationally very expensive. In addition it requires a rather complete characterization 
of the structure and chemical composition of the surface. Working atom-by-atom ab-initio 
methods have to know precisely which atom is sitting where. For plasma-exposed surfaces 
this information is not available in most cases. It is thus better not to rely on it at all 
and--following the second approach--to construct instead an effective Hamiltonian for 
that part of the electronic structure which is expected to be mostly involved in the 
collision process. Physical considerations may then be invoked to parameterize 
the model by a few quantities which are easily available and at the same time have a 
clear physical meaning. 

The particular approach we employ is based on an Anderson-Newns-type effective 
Hamiltonian. Following Gadzuk~\cite{Gadzuk67a,Gadzuk67b} it uses classical image 
charges to mimic the long-range exchange interactions (polarization interactions) and a 
multichannel scattering theory to account in the matrix elements for single-electron transfer
for the non-orthogonality of the target and projectile wavefunctions. The non-orthogonality 
of the wavefunctions is also an issue in the Auger channels~\cite{VGB05}. Taking it into account 
makes however the calculation of the Auger matrix elements even more complicated 
than it already is. In the model presented below we ignore therefore in the Auger 
matrix elements the non-orthogonality of the wavefunctions assuming implicitly that
it is less important than the tunneling of the metal wavefunction through the 
potential barrier, arising from the overlap of the ion and surface potentials, which we 
take into account. The good agreement of the rate we get for Auger neutralization 
with the rate given by Wang and coworkers~\cite{WGM01} as well as with the rate 
obtained by an approach based in part on first principles~\cite{VGB05} supports this 
assumption. What speaks against it is the too large ion survival probability we obtain
for large angles of incidence. But the reason for this is most probably the neglect of 
single-electron transfer from deeper lying levels of the surface (core levels) to 
the $\mathrm{1s}$ shell. It is beyond the scope of the present work to include this 
process as well. 

To furnish the formalism with wavefunctions, simple models are used 
for the surface potential and the electronic structure of the projectile, parameterized 
however such that it reproduces measured ionization energies 
and electron affinities. From our previous work on the de-excitation of metastable nitrogen 
molecules on surfaces~\cite{MBF11,MBF12a,MBF12b} and the neutralization of alkaline-earth 
ions on gold surfaces~\cite{PBF15a,PBF15b} we expect this type of modeling to provide also 
reasonable matrix elements for the Anderson-Newns Hamiltonian describing ion-induced electron 
ejection from metal surfaces.

\subsection{Electronic configurations and energy levels}

To analyze the chain of processes outlined in Fig.~\ref{outlineProcesses} we consider the 
following electronic configurations for the He projectile: ${\rm He}^+{\rm (1s)}$, 
${\rm He}^0{\rm (1s^2)}$, ${\rm He}^*{\rm (1s2s)}$, and ${\rm He}^{*-}{\rm (1s2s^2)}$. 
Without loss of generality we assume the electron of the $\mathrm{He}^+{\rm (1s)}$ ion to 
have spin up. This leaves us with two non-degenerate metastable levels $\mathrm{He}^\ast{\rm (1s2s)}$, 
a triplet ${\rm 2}^3{\rm S}_1$ and a singlet ${\rm 2}^1{\rm S}_0$ with, respectively, a spin-up 
and spin-down electron in the $2s$ shell. The term symbols for the positive ion and the 
groundstate atom are $1^2{\rm S}_{1/2}$ and $1^1{\rm S}_{0}$. We also consider the negative 
ion $\mathrm{He}^{\ast-}{\rm (1s2s^2)}$ arising from either one of the metastable states. 
In both cases the term symbol is $2^2{\rm S}_{1/2}$ because the two electrons in the $\mathrm{2s}$ 
shell have antiparallel spin. It is the lowest lying negatively charged state and known to 
act as an intermediary in surface-induced spin-flip collisions~\cite{HC91,BTG94}. 
It may thus also play a role in secondary electron emission. In principle there are of 
course additional configurations possible. For instance, the metastable state 
${\rm He}^*{\rm (1s2p)}$ could be also involved. We expect it however to be less important 
for the collision we consider because $\mathrm{p}$ orbitals lead to smaller matrix elements 
and thus to smaller transition rates.

Far away from the surface the projectile configurations are characterized by a discrete 
set of energies representing the ionization energies or electron affinities depending
on whether the configurations are electrically neutral, positive, or negative. For the 
reaction scheme shown in Fig.~\ref{outlineProcesses} we need the single-electron 
ionization energies $\mathcal{I}_\HeliumGroundState$, $\mathcal{I}_\HeliumMetaSing$, 
and $\mathcal{I}_\HeliumMetaTrip$, that is, the thresholds of the first ionization
continua of the helium configurations given in the subscripts, as well as the single-electron 
affinities $\mathcal{A}_\HeliumMetaSing$ and $\mathcal{A}_\HeliumMetaTrip$, where the subscripts
indicate again the configurations the energies belong to. How these (positive) energies 
relate to the vacuum level is shown in Fig.~\ref{physicalProcesses} together with 
the processes they are involved in. While the projectile approaches the surface the 
energy levels shift. Assuming a polarization-induced image charge interaction to be
responsible for the shifts, the ionization levels move upward in energy whereas affinity 
levels move downwards~\cite{NMB83}. Close to the surface short-range interactions may
modify the shifts~\cite{MMM98}. The processes we are interested in occur however sufficiently 
far away from the surface that short-range interactions are not yet important. To take all 
this into account, we define five time-dependent single-electron energy levels,
\begin{align}
        \varepsilon^0_{1s\downarrow}(t) &= -\mathcal{I}_\HeliumGroundState + \dfrac{e^2}{4(z(t) - z_i)}\,,
        \label{1sShift}\\
        \varepsilon^\ast_{2s\downarrow}(t) &= -\mathcal{I}_\HeliumMetaSing + \dfrac{e^2}{4(z(t) - z_i)}\,,
        \label{2sSshift}\\
        \varepsilon^\ast_{2s\uparrow}(t) &= -\mathcal{I}_\HeliumMetaTrip + \dfrac{e^2}{4(z(t) - z_i)}\,,
        \label{2sTshift}\\
        \varepsilon^-_{2s\downarrow}(t) &= -\mathcal{A}_\HeliumMetaSing - \dfrac{e^2}{4(z(t) - z_i)}\,,
        \label{2sSNshift}\\
        \varepsilon^-_{2s\uparrow}(t) &= -\mathcal{A}_\HeliumMetaTrip - \dfrac{e^2}{4(z(t) - z_i)}\,,
        \label{2sTNshift}
\end{align}
with the subscript indicating the shell and the spin of the electron and the time-dependence 
arising from the collision trajectory, 
\begin{align}
        z(t)=z_{\rm TP} + v_{\perp} \vert t\vert\,,
\label{trajectory}
\end{align}
where $v_\perp$ is the projectile's velocity component perpendicular to the surface and $z_{\rm TP}$ 
is the turning point of the trajectory. The energy levels~\eqref{1sShift}--\eqref{2sTNshift} 
are thus time-dependent ionization energies and electron affinities.
Note, $z(t)$ describes the classical center-of-mass 
motion of the projectile resulting from the trajectory approximation~\cite{Modinos87} being justified 
because of the large mass of the projectile. The turning point $z_{\rm TP}$ is usually a few Bohr 
radii before the crystallographic ending of the surface. It arises from short-range 
repulsive forces. Our choice for $z_{\rm TP}$, which in general depends on the 
projectile and the target, is guided by the calculations of Lancaster and 
coworkers~\cite{LKW03} showing that the neutralization of $\mathrm{He}^+$ ions at impact 
energies $E_{{\rm kin}\perp} < 60~\mathrm{eV}$, which is also the upper limit in the gracing 
incident experiments we compare our results with, takes typically place $2-5\,\mathrm{a_B}$ 
in front of the surface where $\mathrm{a_B}$ denotes the Bohr radius. We can thus choose 
$z_{\rm TP}=2.27\,\mathrm{a_B}$, as suggested by Modinos and Easa~\cite{ME87}, without 
affecting the charge-transfer too much. Indeed our final results  
are rather robust against changes in $z_{\rm TP}$ up to $\pm \mathrm{a_B/2}$.  
The position of the image plane 
$z_i$, appearing in~\eqref{1sShift}--\eqref{2sTNshift}, is used as a fitting parameter 
but it should be around $1-2\,\mathrm{a_B}$~\cite{LK70}. 
%Our calculations show however that 
%the electron emission spectrum depends only slightly on the position of the image plane. }

In addition to the energy levels of the projectile we also need the energy 
$\varepsilon_{\vec{k}\sigma}$ of an electron in the conduction band of the metal and the 
energy $\varepsilon_{\vec{q}\sigma}(t)$ of an unbound electron at position $z(t)$ in front
of the surface. Modeling, as in our previous work~\cite{MBF11,MBF12a,MBF12b,PBF15a,PBF15b},
the metal by a three-dimensional step potential,
\begin{align}
V_{\mathrm{S}}(z)=-V_0\theta(-z) 
\label{StepPot}
\end{align}
with depth $V_0=E_{\rm F}+\phi$, where 
$E_{\rm F}>0$ is the Fermi energy of the metal and $\phi>0$ the work function,
\begin{align}
        \varepsilon_{\vec{k}\sigma} = \frac{\hbar^2\vec{k}^2}{2m^*_e} - V_0
        \label{CBshift}
\end{align}
with $m^*_e$ the effective mass of an electron in the conduction band of the metal. 
Assuming moreover a plane wave for the wavefunction of an unbound electron in front
of the surface, its energy is given by
\begin{align}
        \varepsilon_{\vec{q}\sigma}(t) = \frac{\hbar^2\vec{q}^{\,2}}{2m_e} - \dfrac{e^2}{4(z(t) - z_i)}\,,
        \label{Cshift}
\end{align}
where the second term takes the interaction of the electron with its image into account.
 
\begin{figure}[t]
        \includegraphics[width=0.95\linewidth]{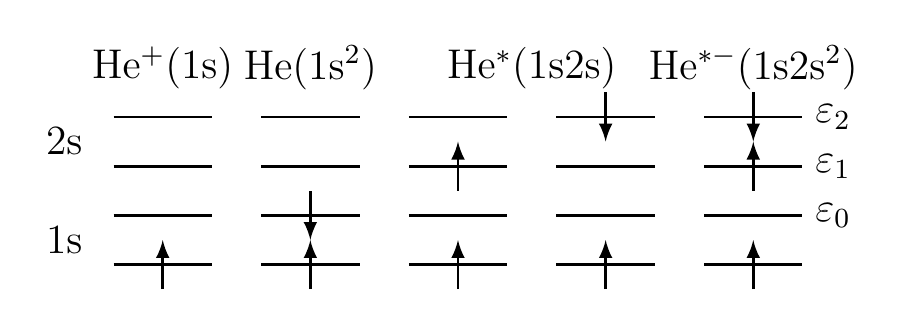}
        \caption{Electronic configurations of the $\mathrm{He}$ projectile included into 
        our modeling. As indicated on the left, the lowest two levels stand for the $\mathrm{1s}$ 
        shell and the upper two for the $\mathrm{2s}$ shell. Since the $\mathrm{1s}$ shell is 
        by assumption always occupied by a spin-up electron, only the energy levels  
        $\varepsilon_0$, $\varepsilon_1$, and $\varepsilon_2$ are of interest. The groundstate
        arises if $\varepsilon_0$ is occupied by a spin-down electron, the metastable triplet
        (singlet) state if $\varepsilon_1$ ($\varepsilon_2$) is occupied by a spin-up (spin-down) 
        electron, and the negative ion state if $\varepsilon_1$ and $\varepsilon_2$ are, 
        respectively, occupied by a spin-up and spin-down electron. Depending on the occupancy
        and the way it is realized the energy levels $\varepsilon_1$ and $\varepsilon_2$ 
        take on different numerical values. This can be organized with projection operators. 
        Two auxiliary bosons finally allow to switch as required between the configurations.}
        \label{configurations}
\end{figure}

Since by assumption the $\mathrm{1s}$ shell is always occupied by a spin-up electron, we in effect 
model the projectile by a three-level system with energies $\varepsilon_0$, $\varepsilon_1$, and 
$\varepsilon_2$ as illustrated in Fig.~\ref{configurations}. An important feature of the model
is that the energies depend on the occupancy of the levels and--in the case of the negative
ion configurations--on the way the occupancy was build-up. To take this into account we 
employ operators
\begin{align}
\label{projectionOperator}
P_{n_0n_1n_2} = \vert n_0n_1n_2 \rangle \langle n_0n_1n_2 \vert \,
\end{align}
projecting onto states $\vert n_0n_1n_2 \rangle$ of the three-level system containing 
$n_i=0, 1$ electrons in the energy levels $\varepsilon_i$. Defining   
\begin{align}
\label{projectionE1sp}
P_{100} \varepsilon_0(t) &= \varepsilon^0_{1s\downarrow}(t)\,,\\
\label{projectionE2spp}
P_{010} \varepsilon_1(t) &= \varepsilon^\ast_{2s\uparrow}(t)\,,\\
\label{projectionE2smp}
P_{001} \varepsilon_2(t) &= \varepsilon^\ast_{2s\downarrow}(t)\,,\\
\label{projectionE2spm}
P_{011} \varepsilon_1(t) &= \varepsilon^{-}_{2s\uparrow}(t)\,,\\
\label{projectionE2smm}
P_{011} \varepsilon_2(t) &= \varepsilon^{-}_{2s\downarrow}(t)\,
\end{align}
with projections to the remaining states required for the completeness
\begin{align}
Q=\sum_{n_0}\sum_{n_1} \sum_{n_2}\vert n_0 n_1 n_2 \rangle \langle  n_0 n_1 n_2 \vert = 1
\label{completeness}
\end{align}
to be zero, it is possible to adjust $\varepsilon_0$, $\varepsilon_1$, and $\varepsilon_2$ 
to the internal energetics of the projectile configurations involved in the 
atom-surface collision we want to model. The operator $Q$ defined in~\eqref{completeness}
will be also required in the quantum kinetic approach described in the section~\ref{QuantumKinetics}. 

\subsection{Wave functions and matrix elements}

To set up the Anderson-Newns Hamiltonian for the charge-transferring atom-surface collision 
processes we are interested in we require a series of matrix elements. Their calculation is 
based on a particular choice of wavefunctions which we now describe. 

As in our previous work~\cite{PBF15a}, the 
electronic states of the metal are the wavefunctions $\psi_{\vec{k}\sigma}(\vec{r}\,)$ of 
the step potential~\eqref{StepPot}. For $k_z<\sqrt{2m^*_eV_0/\hbar^2}$ they describe 
bound electrons whereas for $k_z>\sqrt{2m^*_eV_0/\hbar^2}$ they contain a transmitted and a
reflected wave. From the work of K\"urpick and Thumm~\cite{KT96} we expect little changes
had we used other wavefunctions for the surface based, for instance, on the Jenning-Jones-Weinert 
potential~\cite{JJW88} instead of the potential step. For the states of 
the projectile's $\mathrm{1s}$ and $\mathrm{2s}$ 
shell we take hydrogen wavefunctions $\psi_{1\sigma}(\vec{r}\,)$ and 
$\psi_{2\sigma}(\vec{r}\,)$ with effective charges $Z_{\rm eff}$ adjusted to reproduce the 
ionization energies and electron affinities, $\mathcal{I}_\HeliumGroundState$, 
$\mathcal{I}_\HeliumMetaSing$, $\mathcal{I}_\HeliumMetaTrip$, $\mathcal{A}_\HeliumMetaSing$, 
and $\mathcal{A}_\HeliumMetaTrip$. For the $\mathrm{1s}$ shell the modified hydrogen wavefunction is in 
excellent agreement with the Roothaan-Hartree-Fock $1s$ wavefunction for the helium groundstate 
given by Clementi and Roetti~\cite{CR74}. To estimate the quality of the wavefunction for the 
$\mathrm{2s}$ shell we compared it--due to lack of Roothaan-Hartree-Fock calculations for excited 
helium states--with the Roothaan-Hartree-Fock $2s$ wavefunction of the lithium 
groundstate~\cite{CR74}. As expected, the agreement is not as good as for the $\mathrm{1s}$ 
shell. Since however we found for the metals we investigated charge-transfer to be dominated 
by Auger neutralization, which involves only the $\mathrm{1s}$ shell, we did not attempt to improve 
the wavefunction for the $\mathrm{2s}$ shell. The projectile's continuum states are--as mentioned 
above--approximated by plane waves $\psi_{\vec{q}\sigma}(\vec{r}\,)$. Thereby we ignore 
distortions of the wavefunctions due to the core potential of the projectile, turning plane 
waves into Coulomb waves. It is only an issue for Auger de-excitation and autodetachment, which 
we found however not to be the dominate scattering channels. We did therefore not include this 
complication.

Having wavefunctions we can construct matrix elements for the processes
shown in Fig.~\ref{physicalProcesses}. Denoting the position of the projectile by 
$\vec{r}_p(t) = z(t)\vec{e}_z$ with $z(t)$ defined in~\eqref{trajectory} and following 
Gadzuk~\cite{Gadzuk67a,Gadzuk67b} as well as our earlier work~\cite{MBF12a,PBF15a,PBF15b} we 
obtain
\begin{align}
\label{matrixSET}
V_{\vec{k}\sigma}(t) = \int d^3r\,\psi^\ast_{\vec{k}}(\vec{r}) 
\dfrac{e^2}{\vert \vec{r}-\vec{r}_p(t)\vert} \psi_{2\sigma}(\vec{r}-\vec{r}_p(t))\,
\end{align}
for the matrix element controlling single-electron transfer between the conduction band 
of the surface and the ionization/affinity levels of the projectile originating from 
its $\mathrm{2s}$ shell, 
\begin{align}
\label{matrixAN}
V_{\vec{k}_1\vec{k}_2\vec{k}^\prime\sigma}(t) = \int d^3r&\int d^3r^\prime\, 
\psi^\ast_{1\downarrow}(\vec{r}-\vec{r}_p(t)) \psi^\ast_{\vec{k}^\prime \sigma}(\vec{r}^{\,\prime}) \nonumber\\ 
&\times \dfrac{e^2}{\vert \vec{r}-\vec{r}^{\,\prime}\vert} \psi_{\vec{k}_1 \downarrow}(\vec{r}) 
\psi_{\vec{k}_2 \sigma}(\vec{r}^{\,\prime})\,
\end{align}
for the matrix element driving Auger neutralization into the groundstate, that is, the 
$\mathrm{1s}$ shell of the projectile, and
\begin{align}
\label{matrixDAD}
V_{\vec{k}\vec{k}^\prime\sigma}(t) = \int d^3r&\int d^3r^\prime\, \psi^\ast_{1\downarrow}(\vec{r}-\vec{r}_p(t)) 
\psi^\ast_{\vec{k}^\prime \sigma}(\vec{r}^{\,\prime}) \nonumber\\ 
&\times\dfrac{e^2}{\vert \vec{r}-\vec{r}^{\,\prime}\vert} \psi_{2\downarrow}(\vec{r}-\vec{r}_p(t)) 
\psi_{\vec{k} \sigma}(\vec{r}^{\,\prime})\,,\\
\label{matrixIAD}
V_{\vec{k}\vec{q}\sigma}(t) = \int d^3r&\int d^3r^\prime\, \psi^\ast_{1\downarrow}(\vec{r}-\vec{r}_p(t)) 
\psi^\ast_{\vec{q} \sigma}(\vec{r}^{\,\prime}-\vec{r}_p(t)) \nonumber\\ &
\times\dfrac{e^2}{\vert \vec{r}-\vec{r}^{\,\prime}\vert} \psi_{2\sigma}(\vec{r}^{\,\prime}-\vec{r}_p(t)) 
\psi_{\vec{k} \downarrow}(\vec{r})\,
\end{align}
for the direct and indirect Auger de-excitation, respectively, involving the projectile's $\mathrm{1s}$ 
and $\mathrm{2s}$ shells. Finally, the matrix element for autodetachment reads 
\begin{align}
\label{matrixAuD}
V_{\vec{q}} = \int d^3r\int d^3r^\prime\,
\psi^\ast_{1\downarrow}(\vec{r}) \psi^\ast_{\vec{q} \uparrow}(\vec{r}^{\,\prime})
\dfrac{e^2}{\vert \vec{r}-\vec{r}^{\,\prime}\vert}
\psi_{2\downarrow}(\vec{r}) \psi_{2\uparrow}(\vec{r}^{\,\prime})\,.
\end{align}
In contrast to the other matrix elements it is, within our modeling, independent of time (that is, 
independent of the distance $z(t)$) since it describes a local interaction acting at the 
instantaneous position of the projectile. 

Although the assumptions about the wavefunctions used in~\eqref{matrixSET}--\eqref{matrixAuD} 
are strong, we stick to it because they allow us to pursue the calculation of the matrix 
elements to a large extent analytically by means of lateral Fourier 
transformation, which in turn substantially reduces the numerical effort (which is still large) 
when it comes to the solution of the kinetic equations. To estimate the validity of our approach, 
we compare our results with experimental data. As we will see the agreement is sufficiently good 
to suggest that the approximate matrix elements we use are not too far away from the exact 
matrix elements which we however do not know. Our matrix elements contain a number
of parameters which we list in Table~\ref{tableParameter}. As indicated in the caption of the table 
we use parameters from different sources. If the parameters were given directly for the experiments 
we compare our data with, we took these values. This was the case for the work functions of copper 
and aluminum and for the affinity levels. The rest of the parameters we collected from data tables. 
The effective charge $Z_{\rm eff}$ and the position of the image plane $z_i$ were determined as 
stated above.

\begin{figure}[t]
        \includegraphics[width=0.95\linewidth]{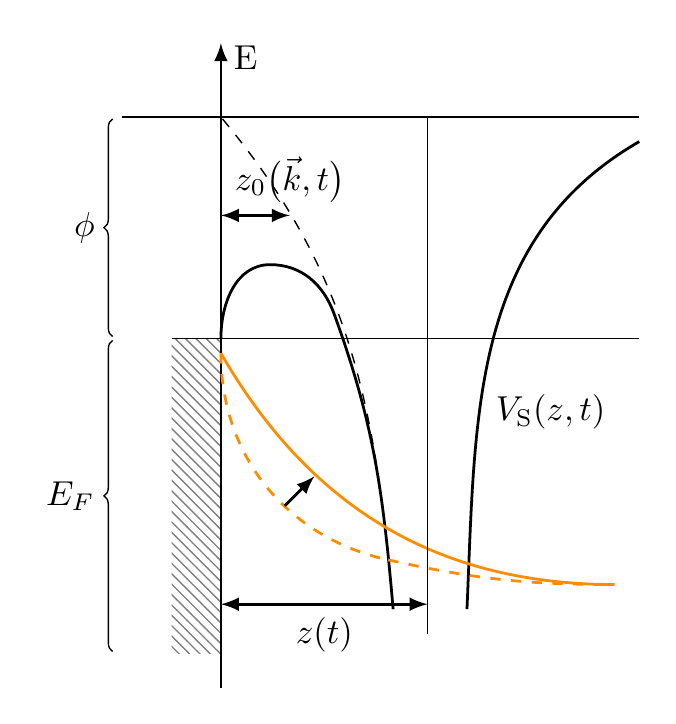}
        \caption{(Color online) Illustration of the ion-induced modification of the step 
         potential used for the surface. The ion, at position $z=z(t)$, creates a potential
         barrier $V_s(z,z(t))$ through which an electron from the conduction band can tunnel. The 
         under-the-barrier motion occurs between the turning points $z=0$ and $z=z_0(\vec{k},t)$. 
         Approximating the latter by the black dashed line simplifies the numerical treatment 
         without loosing accuracy below the Fermi energy $E_{\rm F}$ which is the energy range of 
         interest. The enhancement of the wavefunction for the metal electron filling in 
         an Auger process the projectile's $\mathrm{1s}$ shell at the position of the ion is 
         qualitatively indicated by the solid and dashed orange lines.}
         \label{WKBcartoon}
\end{figure}

An important additional aspect affecting Auger neutralization and indirect Auger de-excitation into 
the projectile groundstate is the enhancement of the wavefunction of the surface electron which 
fills the hole in the $\mathrm{1s}$ shell of the projectile. It arises from the modification of the 
step potential mimicking the surface by the Coulomb potentials of the $\mathrm{He^+}$ ion and its 
image and the image potential of the electron. In effect, the step potential~\eqref{StepPot} 
becomes a potential barrier,
\begin{align}
V_{\mathrm{S}}(z)=V_{\mathrm{S}}(z,z(t))\theta(-z)
\end{align}
with 
\begin{align}
	\label{surfacePotential}
	V_{\mathrm{S}}(z,z(t)) = - \dfrac{e^2}{\vert z(t) - z\vert} 
                                 + \dfrac{e^2}{\vert z(t) + z\vert} - \dfrac{e^2}{4(z-z_i)}~,
\end{align}
as shown in Fig.~\ref{WKBcartoon}, through which the surface electron can tunnel. Following 
Propst~\cite{Propst63} and Penn and Apell~\cite{PA90}, we take this into account by a 
semiclassical correction to the electron's wavefunction using the WKB approximation.
The $z-$dependence of the wavefunction of a metal electron with 
$k_z<\sqrt{2m^*_eV_0/\hbar^2}$ is given by 
\begin{align}
\psi_{\vec{k}\sigma}(z(t)) \propto \exp{\big(\Delta(\varepsilon_{\vec{k}\sigma}, z(t))\big)} \,.
\end{align}
For the step potential,
\begin{align}
\Delta_{\rm step}(\varepsilon_{\vec{k}\sigma}, z(t)) = \kappa_z z(t) 
\label{DeltaStep}
\end{align}
with $\kappa_z = i\sqrt{k_z^2 - 2m^*_eV_0/\hbar^2}$. Using the WKB method to account for the 
tunneling of the electron through the barrier (see Fig.~\ref{WKBcartoon}),
\begin{align}
\Delta_{\rm WKB}(\varepsilon_{\vec{k}\sigma}, z(t)) = 
\int_{0}^{z_0(\varepsilon_{\vec{k}\sigma}, z(t))} K(\varepsilon_{\vec{k}\sigma}, z(t);z)\, dz \,
\end{align}
with $K(\varepsilon_{\vec{k}\sigma}, z(t);z) = i\sqrt{k_z^2 - 2m^*_{e} V_{\mathrm{S}}(z, z(t))/\hbar^2}$
and $z=0$ and $z=z_0(\varepsilon_{\vec{k}\sigma}, z(t))$ the turning points of the under-the-barrier
motion of the electron. Neglecting in~\eqref{surfacePotential} the image potential of the metal electron, 
that is, the third term,
$z_0(\varepsilon_{\vec{k}\sigma}, z(t))$ can be determined analytically leading to the black dashed line 
in Fig.~\ref{WKBcartoon}. By numerical integration we 
then find the adjustment ratio, 
\begin{align}
	\label{adjustmentWKB}
	\Delta_{\rm adjust}(\varepsilon_{\vec{k}\sigma}, z(t)) = 
         \dfrac{ \Delta_{\rm WKB}(\varepsilon_{\vec{k}\sigma}, z(t))}{\Delta_{\rm step}(\varepsilon_{\vec{k}\sigma}, z(t))} 
         \approx 0.15 \sqrt{z(t)} \,,
\end{align}
which depends only weakly on $\vec{k}$. Hence, by replacing $\kappa_z$ in Eq.~\eqref{DeltaStep} by 
\begin{align}
	\label{WKBfinal}
	\kappa_z(t) = \Delta_{\rm adjust}(t) \kappa_z \,,
\end{align}
we approximately take into account the tunneling-induced enhancement of the metal electron wavefunction 
at the projectile position $z(t)$ as illustrated in Fig.~\ref{WKBcartoon}. The assumptions made in the 
calculation of~\eqref{adjustmentWKB} hold as long as $\varepsilon_{\vec{k}\sigma}$ is below the Fermi 
energy $E_{\rm F}$. This is, however, the case since the electron filling the $\mathrm{1s}$ shell
in an Auger process originates from an occupied state of the conduction band of the surface. In 
Section~\ref{Results} we will see that the WKB correction brings the transition rate for Auger 
neutralization we obtain in very good agreement with the rate given by Wang and coworkers~\cite{WGM01}.
It is based on the work of Lorente and Monreal~\cite{LM96} and has been also used by 
others~\cite{LKW03,MMM98}. For the distances we are interested 
in it is moreover in reasonable agreement with calculations based in-part on first 
principles~\cite{VGB05}.

\subsection{Hamiltonian}
We now have everything needed to begin the construction of the Hamiltonian for the processes 
outlined in Fig.~\ref{outlineProcesses}. With the energy shifts and matrix elements given 
above it reads  
\begin{align}
\label{Hamiltonian}
&H(t) = \sum_{n_0n_1n_2} P_{n_0n_1n_2} \big( \varepsilon_0(t) c^\dagger_0 c^\phdag_0 + \varepsilon_1(t) c^\dagger_1 c^\phdag_1 + \varepsilon_2(t) c^\dagger_2 c^\phdag_2\big) \nonumber\\
	&+ \sum_\sigma \omega_\sigma(t) b^\dagger_\sigma b^\phdag_\sigma + \sum_{\vec{k} \sigma} \varepsilon_{\vec{k} \sigma} c^\dagger_{\vec{k} \sigma} c^\phdag_{\vec{k} \sigma} + \sum_{\vec{q} \sigma}\varepsilon_{\vec{q} \sigma}(t) c^\dagger_{\vec{q} \sigma} c^\phdag_{\vec{q} \sigma} \nonumber\\
	&+ \sum_{\vec{k}} \big[ ( P_{000} + P_{010} ) V_{\vec{k}\uparrow}(t) c^\dagger_{\vec{k} \uparrow} c^\phdag_1  + \text{H.c.} \big] \nonumber\\
	&+ \sum_{\vec{k}} \big[ ( P_{000} + P_{001} ) V_{\vec{k}\downarrow}(t) c^\dagger_{\vec{k} \downarrow} c^\phdag_2  + \text{H.c.} \big] \nonumber\\
	&+ \sum_{\vec{k}} \big[ ( P_{001} + P_{011} ) V_{\vec{k}\uparrow}(t) c^\dagger_{\vec{k} \uparrow} b^\dagger_\uparrow c^\phdag_1  + \text{H.c.} \big] \nonumber\\
	&+ \sum_{\vec{k}} \big[ ( P_{010} + P_{011} ) V_{\vec{k}\downarrow}(t) c^\dagger_{\vec{k} \downarrow} b^\dagger_\downarrow c^\phdag_2  + \text{H.c.} \big] \nonumber\\
	&+ \sum_{\vec{k}_1\vec{k}_2\vec{k}^\prime \sigma} \big[ ( P_{000} + P_{100} ) V_{\vec{k}_1\vec{k}_2\vec{k}^\prime \sigma}(t) c^\dagger_{\vec{k}^\prime \sigma} c^\dagger_0 c^\phdag_{\vec{k}_1 \downarrow} c^\phdag_{\vec{k}_2 \sigma}  + \text{H.c.} \big] \nonumber\\
	&+ \sum_{\vec{k}\vec{k}^\prime \sigma} \big[ ( P_{100} + P_{001} ) V_{\vec{k}\vec{k}^\prime \sigma}(t) c^\dagger_{\vec{k}^\prime \sigma} c^\dagger_0 c^\phdag_{\vec{k} \sigma} c^\phdag_2  + \text{H.c.} \big] \nonumber\\
	&+ \sum_{\vec{k}\vec{q}} \big[ ( P_{100} + P_{010} ) V_{\vec{k}\vec{q}\uparrow}(t) c^\dagger_{\vec{q} \uparrow} c^\dagger_0 c^\phdag_{\vec{k} \downarrow} c^\phdag_1  + \text{H.c.} \big] \nonumber\\
	&+ \sum_{\vec{k}\vec{q}} \big[ ( P_{100} + P_{001} ) V_{\vec{k}\vec{q}\downarrow}(t) c^\dagger_{\vec{q} \downarrow} c^\dagger_0 c^\phdag_{\vec{k} \downarrow} c^\phdag_2  + \text{H.c.} \big] \nonumber\\
	&+ \sum_{\vec{q}} \big[ ( P_{100} + P_{011} ) V_{\vec{q}}\, c^\dagger_{\vec{q} \uparrow} c^\dagger_0 c^\phdag_{1} c^\phdag_2  + \text{H.c.} \big]		
	\, ,
\end{align}
where the fermionic operators $c_i^{(\dagger)}$ annihilate (create) an electron in the level $\varepsilon_i$ 
with the spin as indicated in Fig.~\ref{configurations}. Likewise the fermionic operators $c_{\vec{k}\sigma}^{(\dagger)}$
and $c_{\vec{q}\sigma}^{(\dagger)}$ annihilate (create), respectively, an electron with spin $\sigma$ 
in the conduction band of the target surface or the continuum of the projectile. The projection operators 
as defined in~\eqref{projectionOperator} guarantee that each individual term is projected onto 
that subspace of the three level system representing its physical domain of applicability. For instance, 
the term describing Auger neutralization (fifth last term) must contain a factor  $P_{000}+P_{100}$ 
because it involves only the positive ion and the groundstate, that is, in the notation of the three-level
system, the states $\vert 000 \rangle$ and $\vert 100 \rangle$. 

An essential aspect of our approach is that it allows to treat electronic configurations of the 
projectile with defects in their internal energies. More specifically, the numerical value of the 
energy level $\varepsilon_2$ depends on the occupancy of the three-level system. In case $\varepsilon_1$
and $\varepsilon_2$ are occupied, $\varepsilon_2$ denotes an affinity levels of either 
$\HeliumMetaSing$ or $\HeliumMetaTrip$ whereas in the case $\varepsilon_1$ is empty 
$\varepsilon_2$ stands for the ionization level of either $\HeliumMetaSing$ or $\HeliumMetaTrip$. To 
switch between ionization and affinity levels we introduce two auxiliary bosons $b^{(\dagger)}_\uparrow$  
and $b^{(\dagger)}_\downarrow$ with energy  
\begin{align}
\omega_\sigma (t) = \varepsilon^-_{2s-\sigma} (t) - \varepsilon^*_{2s-\sigma} (t)\,,
\end{align}
where $\sigma=\uparrow, \downarrow$ labels the complementary spin orientation of the electron in 
the $\mathrm{2s}$ shell of the two configurations between the boson is expected to switch. With this 
trick~\cite{MBF12b} all processes encoded in the Hamiltonian conserve energy irrespective of whether 
a negative ion or a metastable configuration is involved. 

The Hamiltonian~\eqref{Hamiltonian} is rather involved but the physical meaning of the various terms 
is almost selfexplaining. For instance, the first term describes the ionization and affinity 
levels of the projectile while the next three denote the auxiliary bosons, the continuum of surface states, 
and the continuum of the projectile. The following four terms are the single-electron transfers into 
and out of the metastable ionization and affinity levels. Auger neutralization of the positive 
ion, direct Auger de-excitation of the metastable singlet configuration, indirect Auger de-excitation 
of the metastable triplet and singlet configurations, and the autodetachment of the negative ion 
are given by the last five terms. Note, due to the Pauli principle direct Auger de-excitation is only 
possible for the singlet metastable state (see Fig.~\ref{configurations} and Fig.~\ref{physicalProcesses}). 
Hence, it affects only the levels $\varepsilon_2$ and $\varepsilon_0$. Indirect Auger de-excitation, in 
contrast, is not restricted in such a way.

Working directly with the Hamiltonian~\eqref{Hamiltonian} is cumbersome because it is not suited for a 
diagrammatic analysis which on the other hand is a powerful tool to derive kinetic equations as shown by 
Langreth and coworkers~\cite{LN91,SLN94a,SLN94b}. We rewrite therefore the states making up the projection 
operators in terms of pseudo-particle operators $e^\dagger$, $d^\dagger$ and $s_{n\sigma}^\dagger$ defined by 
\begin{align}
	\label{allPseudoStates}
        \vert000\rangle &= e^\dagger\vert {\rm vac}\rangle \,,\,  
        \vert011\rangle = d^\dagger\vert {\rm vac}\rangle \,,
	\vert100\rangle = s^\dagger_{1\downarrow}\vert {\rm vac}\rangle \,,\, \nonumber\\ 
        \vert010\rangle &= s^\dagger_{2\uparrow}\vert {\rm vac}\rangle \,,\, 
        \vert001\rangle = s^\dagger_{2\downarrow}\vert {\rm vac}\rangle \,.
\end{align}
Hence, $e^\dagger$, $d^\dagger$, $s^\dagger_{1\downarrow}$, $s^\dagger_{2\uparrow}$, and 
$s^\dagger_{2\downarrow}$ create, respectively, the positive ion, the negative ion, the groundstate, 
the triplet metastable state, and the singlet metastable state. The statistics of 
the operators is fixed by the Fermi statistics of the operators $c_i^{(\dagger)}$ and physical 
considerations~\cite{MBF12b}. Since the positive and negative ion represented, respectively, 
by $\vert 000\rangle$ and $\vert 011\rangle$ contain an odd number of electrons, because of the 
spin-up electron always present in the $\mathrm{1s}$ shell, but not explicitly included in the three 
level system (see Fig.~\ref{configurations}), the operators $e^{(\dagger)}$ and $d^{(\dagger)}$ 
should be endorsed with Fermi statistics. The groundstate and the metastable configurations, 
$\vert 100\rangle$, $\vert 010\rangle$, and $\vert 001\rangle$, on the other hand, carry an even 
number of electrons. Hence, it is natural to endorse the operators 
$s^{(\dagger)}_{1\downarrow}$, $s^{(\dagger)}_{2\uparrow}$, and $s^{(\dagger)}_{2\downarrow}$ 
with Bose statistics. 

The relation between the operators $c_0$, $c_1$, and $c_2$, which are single-electron operators,
and the pseudo-particle operators defined in~\eqref{allPseudoStates}, which create many-electron
states, that is, whole electronic configurations, is found by letting the former act on the 
completeness relation~\eqref{completeness}. The result is
\begin{align}
\label{c0m}
c_0 = c_0 \ast 1 &= \vert 000 \rangle\langle 100 \vert - \vert 010 \rangle\langle 110 \vert \nonumber\\ 
&- \vert 001 \rangle\langle 101 \vert + \vert 011 \rangle\langle 111 \vert\,,\\
\label{c0p}
c^\dagger_0 = c^\dagger_0 \ast 1 &= \vert 100 \rangle\langle 000 \vert - \vert 110 \rangle\langle 010 \vert \nonumber\\ 
&- \vert 101 \rangle\langle 001 \vert + \vert 111 \rangle\langle 011 \vert  \,,\\
\label{c1m}
c_1 = c_1 \ast 1 &= \vert 000 \rangle\langle 010 \vert + \vert 100 \rangle\langle 110 \vert \nonumber\\ &- \vert 001 \rangle\langle 011 \vert - \vert 101 \rangle\langle 111 \vert\,,\\
\label{c1p}
c^\dagger_1 = c^\dagger_1 \ast 1 &= \vert 010 \rangle\langle 000 \vert + \vert 110 \rangle\langle 100 \vert \nonumber\\ 
&- \vert 011 \rangle\langle 001 \vert - \vert 111 \rangle\langle 101 \vert  \,,\\
\label{c2m}
c_2 = c_2 \ast 1 &= \vert 000 \rangle\langle 001 \vert + \vert 100 \rangle\langle 101 \vert \nonumber\\ 
&+ \vert 010 \rangle\langle 011 \vert + \vert 110 \rangle\langle 111 \vert\,,\\
\label{c2p}
c^\dagger_2 = c^\dagger_2 \ast 1 &= \vert 001 \rangle\langle 000 \vert + \vert 101 \rangle\langle 100 \vert \nonumber\\ 
&+ \vert 011 \rangle\langle 010 \vert + \vert 111 \rangle\langle 110 \vert  \,,
\end{align}
where the minus signs guarantee the fulfilment of the anti-commutation relations. 
Inserting~\eqref{c0m}-\eqref{c2p} into~\eqref{Hamiltonian}, carrying out all projections,
and making at the end replacements of the sort 
$\vert 000\rangle\langle 100 \vert \rightarrow e^\dagger s_{1\downarrow}$  
we finally obtain the Anderson-Newns Hamiltonian in pseudo-particle representation: 
\begin{align}
       \label{pseudoHamiltonian}
       H(t) &= \varepsilon^0_{1s\downarrow}(t) s^\dagger_{1\downarrow} s^\phdag_{1\downarrow}
       + \sum_\sigma \varepsilon^*_{2s\sigma}(t) s^\dagger_{2\sigma} s^\phdag_{2\sigma} \nonumber\\
       &+ \big[\varepsilon^-_{2s\uparrow}(t) + \varepsilon^-_{2s\downarrow}(t)\big]
       d^\dagger d^\phdag
       + \sum_\sigma \omega_\sigma(t) b^\dagger_\sigma b^\phdag_\sigma \nonumber\\
       &+ \sum_{\vec{k} \sigma} \varepsilon_{\vec{k} \sigma}
         c^\dagger_{\vec{k} \sigma} c^\phdag_{\vec{k} \sigma} 
       + \sum_{\vec{q} \sigma} \varepsilon_{\vec{q}\sigma}(t)
        c^\dagger_{\vec{q} \sigma} c^\phdag_{\vec{q} \sigma} \nonumber\\
       &+ \sum_{\vec{k} \sigma} \big[ V_{\vec{k}\sigma}(t)
        c^\dagger_{\vec{k} \sigma} e^\dagger s^\phdag_{2\sigma} + \text{H.c.} \big] \nonumber\\
       &- \sum_{\vec{k} \sigma} \big[ \sgn(\sigma) V_{\vec{k}\sigma}(t) c^\dagger_{\vec{k} \sigma}
        b^\dagger_\sigma s^\dagger_{2-\sigma} d^\phdag + \text{H.c.} \big] \nonumber\\
       &+ \sum_{\vec{k}_1\vec{k}_2\vec{k}^\prime \sigma} \big[ V_{\vec{k}_1\vec{k}_2\vec{k}^\prime \sigma}(t)
        c^\dagger_{\vec{k}^\prime \sigma} s^\dagger_{1\downarrow} e^\phdag c^\phdag_{\vec{k}_1 \downarrow}
        c^\phdag_{\vec{k}_2 \sigma}  + \text{H.c.} \big] \nonumber\\
       &+ \sum_{\vec{k}\vec{k}^\prime \sigma} \big[ V_{\vec{k}\vec{k}^\prime \sigma}(t)
         c^\dagger_{\vec{k}^\prime \sigma} s^\dagger_{1\downarrow}
         c^\phdag_{\vec{k} \sigma} s^\phdag_{2\downarrow} + \text{H.c.} \big] \nonumber\\
        &+ \sum_{\vec{k}\vec{q}\sigma} \big[ V_{\vec{k}\vec{q}\sigma}(t) c^\dagger_{\vec{q} \sigma}
         s^\dagger_{1\downarrow} c^\phdag_{\vec{k} \downarrow} s^\phdag_{2\sigma}  + \text{H.c.} \big] \nonumber\\
       &+ \sum_{\vec{q}} \big[ V_{\vec{q}}\, c^\dagger_{\vec{q} \uparrow} s^\dagger_{1\downarrow} d^\phdag + \text{H.c.} \big] \,.
\end{align}
The physical meaning of the various terms of the Hamiltonian is now particularly transparent. 
Consider, for instance, the fourth last term. It describes Auger neutralization (and its reverse which 
has to be included to make the Hamiltonian Hermitian) and hence the creation/annihilation of the 
projectile groundstate and a secondary electron by simultaneously annihilating/creating a positive ion 
and two metal electrons. Likewise the last term describes autodetachment (and its reverse), that is, 
the creation/annihilation of the groundstate by annihilation/creation of the negative ion due to 
creating/annihilating an electron in the continuum of the projectile.  In the next section we will
use this Hamiltonian to determine the probabilities with which the various projectile configurations 
appear and an electron is emitted in the course of the atom-surface collision.

\section{Quantum kinetics}
\label{QuantumKinetics}

\begin{figure}[t]
        \includegraphics[width=0.95\linewidth]{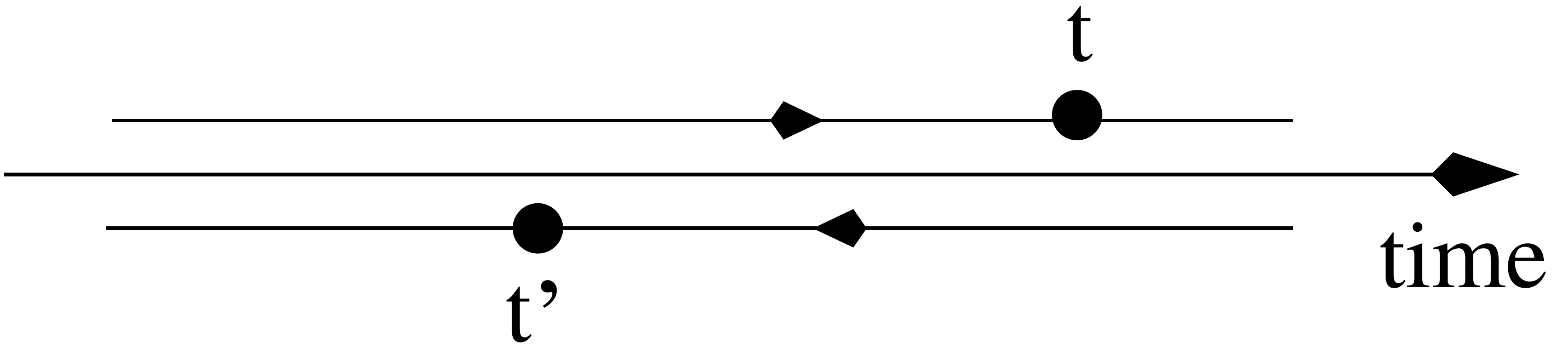}
        \caption{Keldysh contour in the complex time domain running infinitesimally
         above and below the real-time axis from $t=-\infty$ to $t=+\infty$ and 
         back. The time variables $t$ and $t^\prime$ of the contour-ordered Green 
         functions~\eqref{GFe}--\eqref{Baux} and the selfenergies associated with 
         them vary along this contour. To obtain the analytic pieces of these 
         functions--denoted by less-than, greater-than, and retarded--an analytic 
         continuation to the real-time axis is performed after the time-ordering along 
         the contour has been taken into account. This is equivalent to Keldysh's 
         matrix notation for the Green functions~\cite{Keldysh65}.}
        \label{KeldyshContour}
\end{figure}

With the electronic configurations of the He projectile encoded in an effective three-level 
system holding either none, one, or two electrons with the spin polarizations 
given in Fig.~\ref{configurations} we can now calculate the probability with which an electron
is emitted via Auger neutralization or the sequence of single-electron transfers leading to 
Auger de-excitation or autodetachment as shown in Fig.~\ref{outlineProcesses}. For that 
purpose we use the quantum-kinetic method which rests in our case on the contour-ordered 
Green functions~\cite{KB62,Keldysh65},
\begin{align}
iE(t,t^\prime) &= \bigl\langle T_\mathcal{C} \, e(t) \, e^\dag(t^\prime) \bigr\rangle \;, \label{GFe}\\
iS_{1\downarrow}(t,t^\prime) &= \bigl\langle T_\mathcal{C} \, s_{1\downarrow}(t) \, 
s_{1\downarrow}^\dag(t^\prime) \bigr\rangle \;, \\
iS_{2\sigma}(t,t^\prime) &= \bigl\langle T_\mathcal{C} \, s_{2\sigma}(t) \, 
s_{2\sigma}^\dag(t^\prime) \bigr\rangle \;, \\
iD(t,t^\prime) &= \bigl\langle T_\mathcal{C} \,
d(t) \, d^\dag(t^\prime) \bigr\rangle \;,\\
iG_{\vec{q}\sigma}(t,t^\prime) &= \bigl\langle T_\mathcal{C} \,
c_{\vec{q}\sigma}(t) \, c_{\vec{q}\sigma}^\dag(t^\prime) \bigr\rangle\,,\\ 
iG_{\vec{k}\sigma}(t,t^\prime) &= \bigl\langle T_\mathcal{C} \,
c_{\vec{k}\sigma}(t) \, c_{\vec{k}\sigma}^\dag(t^\prime) \bigr\rangle \,, \label{GFm}\\
iB_{\sigma}(t,t^\prime) &= \bigl\langle T_\mathcal{C} \,
b_{\sigma}(t) \, b_{\sigma}^\dag(t^\prime) \bigr\rangle \,, \label{Baux}
\end{align}
where the time variables run over the Keldysh contour shown in Fig.~\ref{KeldyshContour}.  
The first four functions describe the positive ion, the groundstate, the two metastable 
states, and the negative ion while the last three apply, respectively, to an unbound electron 
in the projectile's continuum, the electrons in the conduction band of the target surface,
and the auxiliary bosons. The operators making up the Green functions evolve in time with the 
full Hamiltonian~\eqref{pseudoHamiltonian} and the brackets denote the statistical average with 
respect to the initial density matrix describing one-auxiliary-boson states, surface 
electrons in thermal equilibrium, and an empty projectile, that is, a positive ion.  

Following the work of Langreth and coworkers~\cite{LN91,SLN94a,SLN94b} we use Dyson equations for 
these functions to derive a set of equations for the occurrence probabilities/occupancies of the 
bound projectile states, that is, the affinity and ionization levels. Introducing the selfenergies 
$\Pi_e(t,t^\prime)$, $\Sigma_{1\downarrow}(t,t^\prime)$, $\Sigma_{2\sigma}(t,t')$, and 
$\Pi_d(t,t^\prime)$ for the Green functions $iE(t,t^\prime)$, $iS_{1\downarrow}(t,t^\prime)$, 
$iS_{2\sigma}(t,t^\prime)$, and $iD(t,t^\prime)$, we obtain ($\hbar=1$ in this section and the 
two appendices)
\begin{align}
\label{rateEqEa}
\frac{d}{dt}n_+(t) &= 2\text{Im}\int_{-\infty}^\infty d\bar{t} E^<(\bar{t},t)\Pi_e^R(t,\bar{t})  \nonumber\\
                   &- 2\text{Im}\int_{-\infty}^\infty d\bar{t} E^R(t,\bar{t})\Pi_e^<(\bar{t},t)~, \\
\label{rateEqSa1}
\frac{d}{dt}n_g(t) &= 2\text{Im}\int_{-\infty}^\infty d\bar{t} 
                      S^<_{1\downarrow}(\bar{t},t)\Sigma_{1\downarrow}^R(t,\bar{t})  \nonumber\\
                   &- 2\text{Im}\int_{-\infty}^\infty d\bar{t} S^R_{1\downarrow}(t,\bar{t})\Sigma_{1\downarrow}^<(\bar{t},t)~, \\
\label{rateEqSa2}
\frac{d}{dt}n_{\sigma}(t) &= 2\text{Im}\int_{-\infty}^\infty d\bar{t} 
                      S^<_{2\sigma}(\bar{t},t)\Sigma_{2\sigma}^R(t,\bar{t})  \nonumber\\
                   &- 2\text{Im}\int_{-\infty}^\infty d\bar{t} S^R_{2\sigma}(t,\bar{t})\Sigma_{2\sigma}^<(\bar{t},t)~, \\
\label{rateEqDa}
\frac{d}{dt}n_-(t) &= 2\text{Im}\int_{-\infty}^\infty d\bar{t} 
                    D^<(\bar{t},t)\Pi_d^R(t,\bar{t})  \nonumber\\
                   &- 2\text{Im}\int_{-\infty}^\infty d\bar{t} D^R(t,\bar{t}) \Pi_d^<(\bar{t},t)~, 
\end{align}
for the time evolution of the probabilities $n_e(t)$, $n_g(t)$, $n_\sigma(t)$, and 
$n_-(t)$ with which, respectively, the positive ion, the groundstate, the two metastable states
($\sigma=\uparrow$ denoting the triplet and $\sigma=\downarrow$ the singlet), and the negative 
ion occur.
\begin{figure}[t]
        \includegraphics[width=0.95\linewidth]{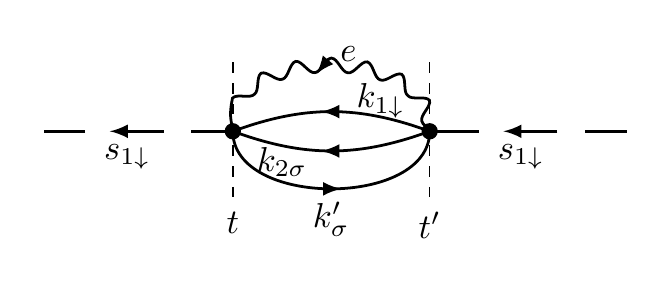}
	\caption{Diagrammatic representation of the selfenergy due to Auger neutralization,
        $\Sigma^{\rm AN}_{1s\downarrow}(t,t')$, entering the Dyson equation for the groundstate 
        propagator $iS_{1\downarrow}(t,t^\prime)$ in the non-crossing approximation. The 
        dashed and wavy lines are the dressed propagators of the groundstate and positive ion, 
        whereas the solid lines mark the undressed propagators for surface electrons. Standard
        diagrammatic rules~\cite{LL81} can be applied to this type of diagrams if one keeps in
        mind that time integrations/variables run over the Keldysh contour~\cite{KB62,Keldysh65}. 
        Details can be found in our previous work~\cite{MBF11,MBF12a,MBF12b,PBF15a}.}
	\label{diagram}
\end{figure}

Equation~\eqref{rateEqEa}--\eqref{rateEqDa} are exact but of course not closed in terms 
of the occurrence probabilities. To proceed  we set up the selfenergies in the non-crossing
approximation, utilize that the matrix elements~\eqref{matrixSET}--\eqref{matrixAuD} factorize 
approximately in functions of $t$ and the set of $\vec{k}$ vectors, and finally apply the 
semiclassical approximation to~\eqref{rateEqEa}--\eqref{rateEqDa} developed by Langreth and 
coworkers~\cite{LN91,SLN94a,SLN94b} which in essence is a saddle-point integration in time.

In order to get an impression about how the selfenergies look like, we show in 
Fig.~\ref{diagram} the contribution to the selfenergy $\Sigma_{1s\downarrow}(t,t')$,
entering the Dyson equation of the groundstate propagator $iS_{1\downarrow}(t,t^\prime)$,
which arises from the Auger neutralization. There are also contributions to 
$\Sigma_{1s\downarrow}(t,t')$ due to direct and indirect Auger de-excitation as well as 
autodetachment. They are given in Appendix~\ref{AppSelf} together with the other
selfenergies entering Eqs.~\eqref{rateEqEa}--\eqref{rateEqDa} and some details concerning 
their calculation. Using standard diagrammatic rules~\cite{LL81} the diagram shown in 
Fig.~\ref{diagram} translates to 
\begin{align}
	-i\Sigma^{\rm AN}_{1s\downarrow}(t,t') = -(i)^2 \sum_{\vec{k}_1 \vec{k}_2 \vec{k}'\sigma}
         V_{\vec{k}_1\vec{k}_2\vec{k}' \sigma}(t) V^\ast_{\vec{k}_1\vec{k}_2\vec{k}' \sigma}(t') \nonumber\\
	\times i E(t,t') i G_{\vec{k}_1\downarrow}(t,t') i G_{\vec{k}_2\sigma} (t,t') i G_{\vec{k}'\sigma} (t',t) \,.
        \label{SelfAuger}
\end{align}
The time variables run over the (Keldysh) contour, that is, from $t=-\infty$ to $t=+\infty$ and back
as shown in Fig.~\ref{KeldyshContour}. Application of the Langreth-Wilkins rules~\cite{LW72} with a subsequent 
projection to the subspace encoded in the completeness relation~\eqref{completeness} yields the analytic 
pieces of the selfenergies, where the time variables are now taken from the real-time axis,
\begin{align}
	\label{analyticContinuation}
	\Sigma_{1\downarrow}^{{\rm AN},\gtrless}(t,t^\prime) &= \sum_{\vec{k}_1\vec{k}_2\vec{k}^\prime\sigma} V_{\vec{k}_1\vec{k}_2\vec{k}^\prime\sigma}(t) V_{\vec{k}_1\vec{k}_2\vec{k}^\prime\sigma}^\ast(t^\prime) \nonumber\\
	\times& E^\gtrless (t,t^\prime) G^\gtrless_{\vec{k}_1\downarrow}(t,t^\prime) G^\gtrless_{\vec{k}_2\sigma}(t,t^\prime) G^\lessgtr_{\vec{k}^\prime\sigma}(t^\prime,t)\,, \\
	\Sigma_{1\downarrow}^{{\rm AN},R}(t,t^\prime) &= \sum_{\vec{k}_1\vec{k}_2\vec{k}^\prime\sigma} V_{\vec{k}_1\vec{k}_2\vec{k}^\prime\sigma}(t) V_{\vec{k}_1\vec{k}_2\vec{k}^\prime\sigma}^\ast(t^\prime) \nonumber\\
	\times& E^R(t,t^\prime) G^>_{\vec{k}_1\downarrow}(t,t^\prime) G^>_{\vec{k}_2\sigma}(t,t^\prime) G^<_{\vec{k}^\prime\sigma}(t^\prime,t)\,,	
\end{align}
with the superscripts $>,<$, and $R$ indicating the less-than, greater-than, and retarded 
pieces of the selfenergy~\eqref{SelfAuger}. In accordance with the non-crossing approximation 
the surface electrons are propagated by the undressed Green function,
\begin{align}
\label{Gundressed}
G^\gtrless_{\vec{k}\sigma}(t,t')=f^\gtrless(\varepsilon_{\vec{k}\sigma}) 
e^{-i\varepsilon_{\vec{k}\sigma}(t-t')} \, .
\end{align}
Only the propagators applying to the ionization and affinity levels of the projectile,
$iE(t,t^\prime)$, $iS_{1\downarrow}(t,t^\prime)$, $iS_{2\sigma}(t,t^\prime)$, and $iD(t,t^\prime)$
are modified by selfenergies.
 
Due to the approximate factorization of the time and momentum dependencies of the matrix 
elements it is possible to express the selfenergies by functions arising from the 
application of the golden rule to the respective interaction terms in the Hamiltonian. 
%We apply this simplification not only to the selfenergies due to single-electron 
%transfer~\cite{LN91} but also to the selfenergies due to two-electron Coulomb interactions
%leading to the Auger processes in~\eqref{pseudoHamiltonian}. 
The physical meaning of the functions is the one of a (partial) levelwidth. Since they eventually determine 
the rates entering the rate equations for the occurrence probabilities given below, we list 
the functions, however, without derivation which is quite lengthy. An exemplary calculation 
is presented in Appendix~\ref{AppWidth}. 

Single-electron processes are characterized by 
\begin{align}
\label{levelwidth2Sp}
\Gamma_{\varepsilon^*_{2s\sigma}}(t) &= 2\pi \sum_{\vec{k}} \vert V_{\vec{k}\sigma}(t) \vert^2 
\delta( \varepsilon^*_{2s\sigma}(t) - \varepsilon_{\vec{k}\sigma} ) \,, \\
\label{levelwidth2Sm}
\Gamma_{\varepsilon^-_{2s\sigma}}(t) &= 2\pi \sum_{\vec{k}} \vert V_{\vec{k}\sigma}(t) \vert^2 
\delta( \varepsilon^-_{2s\sigma}(t) - \varepsilon_{\vec{k}\sigma} ) \,,
\end{align}
while Auger processes are encoded in 
\begin{align}
\label{levelwidthAN}
\Gamma_{\rm AN}(t) =\,& 2\pi \sum_{\vec{k}_1\vec{k}_2\vec{k}^\prime\sigma} \vert V_{\vec{k}_1\vec{k}_2\vec{k}^\prime\sigma}(t) \vert^2 \,\rho_{\vec{k}_1\vec{k}_2\vec{k}'\sigma}(t)\,,\\
\label{levelwidthDAD}
\Gamma_{\rm DAD\downarrow}(t) =\,& 2\pi \sum_{\vec{k}\vec{k}^\prime\sigma} \vert V_{\vec{k}\vec{k}^\prime\sigma}(t) \vert^2 \,\rho_{\vec{k}\vec{k}'\sigma}(t) \,,\\
\label{levelwidthIAD}
\Gamma_{\rm IAD\sigma}(t) =\,& 2\pi \sum_{\vec{k}\vec{q}} \vert V_{\vec{k}\vec{q}\sigma}(t) 
\vert^2 \,\rho_{\vec{k}\vec{q}\sigma}(t) \,, \\
\label{levelwidthAuD}
\Gamma_{\rm AuD} =\,& 2\pi \sum_{\vec{q}} \vert V_{\vec{q}} \vert^2 \, \rho_{\vec{q}} \,,
\end{align}
with
\begin{align}
\label{rhoAN}
	\rho_{\vec{k}_1\vec{k}_2\vec{k}'\sigma}(t) =&\,  f^<(\varepsilon_{\vec{k}_1\downarrow}) 
f^<(\varepsilon_{\vec{k}_2\sigma}) f^>(\varepsilon_{\vec{k}^\prime\sigma}) \nonumber\\
	\times&\delta (\varepsilon^0_{1s\downarrow}(t) - \varepsilon_{\vec{k}_1\downarrow} - 
\varepsilon_{\vec{k}_2\sigma} + \varepsilon_{\vec{k}^\prime\sigma})\,,\\
\label{rhoDAD}
	\rho_{\vec{k}\vec{k}'\sigma}(t) =&\, f^<(\varepsilon_{\vec{k}\sigma}) f^>(\varepsilon_{\vec{k}'\sigma}) \nonumber\\
	\times&\delta ( \varepsilon^0_{1s\downarrow}(t) -\varepsilon^*_{2s\downarrow}(t) - \varepsilon_{\vec{k}\sigma} + \varepsilon_{\vec{k}^\prime\sigma}) \,,\\
\label{rhoIAD}
	\rho_{\vec{k}\vec{q}\sigma}(t) =&\, f^<(\varepsilon_{\vec{k}\downarrow}) f^>(\varepsilon_{\vec{q}\sigma}(t)) \nonumber\\
	\times& \delta (\varepsilon^0_{1s\downarrow}(t) - \varepsilon^*_{2s\sigma}(t) - \varepsilon_{\vec{k}\downarrow} + \varepsilon_{\vec{q}\sigma}(t) ) \,,\\
\label{rhoAuD}
\rho_{\vec{q}}
=\, g^>(\varepsilon_{\vec{q}\uparrow})& \,\delta\big( \varepsilon^0_{1s\downarrow}(t) - \varepsilon^-_{2s-\sigma}(t) - \varepsilon^*_{2s\sigma}(t) + \varepsilon_{\vec{q}\uparrow}(t) \big) \,,
\end{align}
where on the rhs of the last equation $\sigma=\downarrow$ or $\uparrow$ depending 
on whether the negative ion is formed out of $\HeliumMetaSing$ or $\HeliumMetaTrip$. 
Notice, in contrast to the other levelwidth functions, $\Gamma_{\rm AuD}$ does not 
depend on time since the matrix element $V_{\vec{q}}$ is independent of time and the 
time dependencies of the energies in the delta function contained in $\rho_{\vec{q}}$ 
cancel. 

To get the expressions we used the arguments Langreth and Nordlander~\cite{LN91} 
developed for simplifying selfenergies due to single-electron transfer with the 
exception that in the levelwidths arising from Auger and autodetachment processes the 
distribution functions for the surface and continuum electrons, $f^\gtrless(\varepsilon_{\vec{k}\sigma})$
and $g^>(\varepsilon_{\vec{q}\uparrow})$, are not separated out from the summations in momentum
space. The functions $f^<(\varepsilon)$ and $f^>(\varepsilon)=1-f^<(\varepsilon)$ encode, 
respectively, initially occupied and empty states of the conduction band of the metal. Hence, 
$f^<(\varepsilon)$ is the Fermi-Dirac distribution function at temperature $T_s$ of the
surface. The distribution function for an electron in the continuum of the projectile 
$g^>(\varepsilon_{\vec{q}\uparrow})=1$ for $\varepsilon_{\vec{q}\uparrow} > 0$ and equal 
to zero otherwise.

The saddle-point integration in time utilizes the fact that the Green functions lead to 
selfenergies which are strongly peaked at equal times. In effect, the time variables of the 
projectile Green functions (including the ones entering the selfenergies) on the rhs of 
Eqs.~\eqref{rateEqEa}--\eqref{rateEqDa} are set to equal times once the time integrations
are carried out. Identifying less-than functions at equal times with occurrence 
probabilities/occupancies and realizing that retarded functions at equal times are simply 
equal to unity in the time intervals where they do not vanish, we obtain the rate equations
\begin{widetext}
\begin{align}
\label{rateEq}
\frac{d}{dt} 
\begin{pmatrix} 
n_+ \vspace{1.25mm}\\
n_\uparrow \,\vspace{1.25mm}\\
n_\downarrow \,\vspace{1.25mm}\\
n_- \vspace{1.25mm}\\
n_g\,
\end{pmatrix}
= 
\begin{pmatrix} 
-[\Gamma_\uparrow^<\! +\! \Gamma_\downarrow^<\! +\! \Gamma_{\rm AN}^<] & \hspace*{-5mm}\Gamma_\uparrow^> & \hspace*{-5mm}\Gamma_\downarrow^> & \hspace*{-5mm} 0 & \,\,0 \hspace*{1mm}\vspace{1.25mm}\\
\Gamma_\uparrow^<  & \hspace*{-5mm}-[\Gamma_\uparrow^>\! + \!\Gamma^<_{-,\downarrow} \!+\! \Gamma_{\rm IAD\uparrow}^<] & \hspace*{-5mm} 0 & \hspace*{-5mm}\Gamma^>_{-,\downarrow} & \,\,0 \hspace*{1mm}\vspace{1.25mm}\\
\Gamma_\downarrow^< &  \hspace*{-5mm}0 & \hspace*{-5mm}-[\Gamma_\downarrow^> \!+\! \Gamma^<_{-,\uparrow} \!+\! \Gamma_{\rm IAD\downarrow}^< \!+\! \Gamma_{\rm DAD\downarrow}^<]  & \hspace*{-5mm} \Gamma^>_{-,\uparrow} & \,\,0 \hspace*{1mm}\vspace{1.25mm}\\
0 & \hspace*{-5mm}\Gamma^<_{-,\downarrow} &  \hspace*{-5mm}\Gamma^<_{-,\uparrow} & \hspace*{-5mm} -[\Gamma^>_{-,\uparrow} \!+\! \Gamma^>_{-,\downarrow} \!+\! \Gamma^<_{\rm AuD}] & \,\,0 \hspace*{1mm}\vspace{1.25mm}\\
\Gamma_{\rm AN}^<  & \hspace*{-5mm}\Gamma_{\rm IAD\uparrow}^< & \hspace*{-5mm}\Gamma_{\rm IAD\downarrow}^< \!+\! \Gamma_{\rm DAD\downarrow}^< & \hspace*{-5mm} \Gamma^<_{\rm AuD} & \,\,0 \hspace*{1mm}
\end{pmatrix} 
\cdot 
\begin{pmatrix}
n_+ \vspace{1.25mm}\\
n_\uparrow \,\vspace{1.25mm}\\
n_\downarrow \,\vspace{1.25mm}\\
n_- \vspace{1.25mm}\\
n_g \,
\end{pmatrix}  \,
\end{align}
\end{widetext}
which are, due to the completeness~\eqref{completeness}, subject to the constraint
\begin{align}
\label{constraint}
n_+ + n_\uparrow + n_\downarrow + n_- + n_g = 1\,.
\end{align}
The compliance of~\eqref{constraint} can be easily verified by noting 
\begin{align}
\dfrac{d}{dt} \big( n_+ + n_\uparrow + n_\downarrow + n_- + n_g \big) = 0 \,,
\end{align}
and summing each column of~\eqref{rateEq} which results in the nullifying of the 
rates. Thus, the constraint~\eqref{constraint} is fulfilled. Also note, with 
respect to the diagonal of the coefficient matrix in~\eqref{rateEq}, the entries 
in the lower triangle comprise only less-than rates $\Gamma^<_{\dots}$ whereas 
in the upper triangle only greater-than rates $\Gamma^>_{\dots}$ appear. Besides 
having no entries of greater Auger rates $\Gamma^>_{\rm AN}$, $\Gamma^>_{\rm DAD\downarrow}$, 
$\Gamma^>_{\rm IAD\downarrow}$, $\Gamma^>_{\rm IAD\uparrow}$, and $\Gamma^>_{\rm AuD}$ the 
matrix is symmetric.

For the interpretation and numerical solution of~\eqref{rateEq} we apply to the 
rates the adiabatic approximation. The rates in~\eqref{rateEq} can then be expressed
by the levelwidth functions. For the 
single-electron transfers the adiabatic approximation yields~\cite{LN91}
\begin{align}
\label{staticApprox2Sp}
\Gamma^\gtrless_{\sigma}(t) &= \Gamma_{\varepsilon^*_{2s\sigma}}(t) f^\gtrless (\varepsilon^*_{2s\sigma}(t)) \,, \\
\label{staticApprox2Sm}
\Gamma^\gtrless_{-,\sigma}(t) &= \Gamma_{\varepsilon^-_{2s\sigma}}(t) f^\gtrless (\varepsilon^-_{2s\sigma}(t)) \,
\end{align}
with $\Gamma_{\varepsilon^*_{2s\sigma}}(t)$ and $\Gamma_{\varepsilon^-_{2s\sigma}}(t)$
defined in~\eqref{levelwidth2Sp} and~\eqref{levelwidth2Sm}, respectively. The Auger and 
autodetachment transition rates reduce in the adiabatic approximation simply to 
the levelwidths functions given in Eqs.~\eqref{levelwidthAN}--\eqref{levelwidthAuD}. 
Hence,  
\begin{align}
\label{approxAN}
	\Gamma^<_{\rm AN}(t) &= \Gamma_{\rm AN}(t)\,,\\
\label{approxDAD}
	\Gamma^<_{\rm DAD\downarrow}(t) &= \Gamma_{\rm DAD\downarrow}(t)\,,\\
\label{approxIAD}
	\Gamma^<_{\rm IAD\sigma}(t) &= \Gamma_{\rm IAD\sigma}(t)\,, \\
\label{approxAuD}
	\Gamma^<_{\rm AuD} &= \Gamma_{\rm AuD} \,.
\end{align}
At this point one clearly sees that in the derivation of the levelwidths due to Auger 
and autodetachment processes we did not factorize out the distribution functions as 
it is the case in the derivation of the levelwidths due to single-electron transfers. 
As a result, the distribution functions for the metal electron appear in front of the 
width functions in~\eqref{staticApprox2Sp} and~\eqref{staticApprox2Sm} but not 
in~\eqref{approxAN}--\eqref{approxAuD}, where they are contained in the width functions
themselves. 

A particular characteristic of the adiabatic rates, in contrast to the quantum-kinetic 
rates coming out directly from the saddle-point approximation to~\eqref{rateEqEa}--\eqref{rateEqDa}
as discussed in Appendix~\ref{AppSelf}, 
is that they are positive semidefinite. With the adiabatic rates Eq.~\eqref{rateEq} can 
thus be interpreted straightforwardly: The lower triangle describes the gain of the 
projectile configurations by the processes entering this part of the matrix. In terms of 
Fig.~\ref{outlineProcesses} the lower triangle encodes the transitions from left to right 
and from top to bottom, starting with the positive ion which is the initial configuration. 
The diagonal of the matrix gives the losses of the configurations. In contrast to the lower
triangle, the upper triangle describes indirect gains for the configurations.
It encodes the transitions in Fig.~\ref{outlineProcesses} from right to left. Moving 
from bottom to top is not allowed energetically. In case it was, the last column of 
the matrix would be filled with greater-than Auger rates. 

With the rate equation~\eqref{rateEq} it is now particularly easy to write down a 
differential equation for the probability of emitting a secondary electron. Every 
process outlined in Fig.~\ref{outlineProcesses} that leads to the occurrence of the ground 
state $\HeliumGroundState$ generates an excited electron (see Fig.~\ref{physicalProcesses}).
Thus, the rate equation for the probability to emit a secondary electron at time $t$ 
with energy $\varepsilon$ is,
\begin{align}
\label{secE}
\dfrac{d}{dt} \gamma_e(\varepsilon, t) =&\, n_+(t) \bar{\Gamma}^<_{\rm AN}(\varepsilon,t) 
+ n_\uparrow(t) \bar{\Gamma}^<_{\rm IAD\uparrow}(\varepsilon,t)\nonumber\\ 
&+ n_\downarrow(t) \big[ \bar{\Gamma}^<_{\rm IAD\downarrow}(\varepsilon,t) 
+ \bar{\Gamma}^<_{\rm DAD\downarrow}(\varepsilon,t) \big] \nonumber\\
&+ n_-(t) \bar{\Gamma}^<_{\rm AuD}(\varepsilon,t) \,.
\end{align}
It has the same structure as the rate equation for the groundstate. The spectrally 
resolved rates $\bar{\Gamma}^<_{\dots}(\varepsilon,t)$ entering this equation are  
essentially the ones given in~\eqref{approxAN}--\eqref{approxAuD} except that the 
integration over the magnitude of the wave vector $\vec{k}$ of the excited electron 
is not carried out and that the conditions for escaping from the surface have to 
be taken into account~\cite{Baragiola94}. The reason is the following:  An excited 
electron becomes a secondary electron only if it is also able to escape from the 
location where it is generated. If it is created on-site the projectile due to 
autodetachment or indirect Auger de-excitation the electron has to overcome its 
image potential $V_i(z(t))=e^2/4(z(t)-z_i)$ requiring, in the spirit of the escape 
cone model~\cite{FW76}, $q_z>0$ and
\begin{align}
\label{maxAngle}
\theta < \theta^{\rm max} (\varepsilon) = \arccos{\sqrt{V_i(z(t))/\varepsilon}} \,,
\end{align}
where $\theta$ is the angle between $\vec{q}$ and the outward surface normal. 
The $\vec{q}$-integration in $\Gamma^<_{\rm AuD, IAD}(\varepsilon,t)$ is thus cropped leading
to modified rates which we denote in~\eqref{secE} by $\bar{\Gamma}^<_{\rm AuD, IAD}(\varepsilon,t)$. 
In case, the electron is generated inside the solid surface, that is, by Auger neutralization 
or direct Auger de-excitation, the escape of the electron is also affected by scattering 
processes. Assuming elastic scattering to be most important, the electron arrives 
isotropically at the interface leading to the rates $\bar{\Gamma}^<_{\rm AN, DAD}(\varepsilon,t) 
= \mathcal{T}(\varepsilon) \Gamma^<_{\rm AN, DAD}(\varepsilon,t)$ with
\begin{align}
\label{transmission}
\mathcal{T}(\varepsilon) = \dfrac{1}{2} \bigg( 1 - \sqrt{\dfrac{V_0}{\varepsilon + V_0}} \,\bigg) \,,
\end{align}
the surface transmission function~\cite{Baragiola94}. 

Solving~\eqref{secE}, the energy spectrum of the emitted secondary electron is obtained by
\begin{align}
	\label{specDensity}
	\gamma_e(\varepsilon) =	\gamma_e (\varepsilon, t\rightarrow\infty) \,,
\end{align}
and the probability that an electron gets emitted at all, that is, the secondary electron
emission coefficient ($\gamma-$coefficient) follows by integration over all energies,
\begin{align}
	\label{numSecE}
	\gamma_e = \int d\varepsilon\, \gamma_e (\varepsilon, t\rightarrow\infty) \,.
\end{align}

In order to compare our results with experiments we apply one more modification. Surface
scattering experiments typically occur under conditions of grazing incidence~\cite{Winter02,MHB93}. 
The lateral velocity $v_\parallel$ of the projectile is thus very large. To account in our calculations 
for the smearing of the metal electron's Fermi-Dirac distribution induced by the lateral motion of the 
projectile, in addition to the thermal smearing of the distribution function due to the surface temperature 
$T_s$, we replaced for the numerical calculations in the formulas given above the function $f^<(\varepsilon)$ 
by an angle-averaged velocity-shifted distribution~\cite{SHL03},
\begin{align}
f^<(\varepsilon,v_\parallel) = 
\dfrac{\ln{(1+e^{-\beta(\varepsilon+\phi-\delta)})}-\ln{(1+e^{-\beta(\varepsilon+\phi+\delta)})}}
{2\beta \delta}\, 
\label{VelocitySmearing}
\end{align}
with $\phi$ the work function of the surface, $\beta=1/k_B T_s$, and $\delta=k_{\rm F} v_\parallel$, 
where $k_{\rm F}$ is the surface's Fermi wave number. From the projectile's perspective the velocity 
smearing populates surface states above the Fermi energy thereby potentially strengthening 
charge-transfer processes from the metal to the He metastable states which, due to image shifting, 
turn out to be well above the Fermi energy.

Let us finally say a few words about the numerics we applied. The calculation of the levelwidths 
\eqref{levelwidth2Sp}--\eqref{levelwidthAuD} requires at least a two dimensional integration over 
the solid angle of $\vec{k}$ or $\vec{q}$ and at worst, in the case of Auger neutralization, an 
integration in nine dimensions. In the case of indirect Auger de-excitation, an additional 6-dimensional 
numerical integration must be performed over $\vec{r}$ and $\vec{r}~'$, since the method of lateral 
Fourier transformation, unlike for the other channels, does not lead to an analytic result. The 
integrations are done by a MPI parallelized Monte Carlo Vegas code~\cite{Lepage78}
for a discrete number 
of different 
times. To obtain the matrix elements at times in between we utilized multidimensional-linear interpolation. 
The same strategy was used for the additional integrals of the indirect Auger de-excitation. Because 
of the multidimensionality, using more advanced interpolation methods, e.g. splines, would be 
a difficult undertaking, not necessarily leading to better results. In addition, an interpolation of the 
time-arguments of the rates~\eqref{staticApprox2Sp}--\eqref{approxIAD} is necessary to solve the rate 
equation~\eqref{rateEq}. Here, when interpolating, we take advantage of the fact that the rates are 
almost exponential, which greatly improves the results. To solve the rate equation~\eqref{rateEq}, 
finally, we employed the explicit embedded Runge-Kutta Cash-Karp method also provided by the GNU 
scientific library. We have put importance on a reasonable error propagation resulting in a relative 
numerical error of the calculated occurrence/occupation probabilities of less than $10^{-4}$.

\section{Results}
\label{Results}

\begin{table}[t]
        \begin{center}
                \begin{tabular}{c|c|c|c|c|c|c|c}
                        & $\mathcal{I}$[eV] & $\mathcal{A}$[eV] & $Z_{\rm eff}$ & $z_i$[$\rm{a_B}$] & $\phi$[eV] & $E_{\rm F}$[eV]  & $m_e^*/m_e$ \\\hline
                        {$\rm He(1^1S_0)$} & 24.5875 & -- & 1.68 & -- & -- & -- & -- \\
                        {$\rm He^\ast(2^3S_1)$} & 4.7678 & -- & 1.18 & -- & -- & -- & -- \\
                        {$\rm He^\ast(2^1S_0)$} & 3.9716 & -- & 1.08 & -- & -- & -- & -- \\
                        {$\rm He^{\ast}(2^3S_1)$} & -- & 1.25 & 0.61 & -- & -- & -- & -- \\
                        {$\rm He^{\ast}(2^1S_0)$} & -- & 0.45 & 0.36 & -- & -- & -- & -- \\
                        {\rm W(110)} & -- & -- & -- & 1.3 & 5.22 & 6.4 & 1.1 \\
                        {\rm Cu(100)} & -- & -- & -- & 1.3 & 5.1 & 7 & 1.1 \\
                        {\rm Al(100)} & -- & -- & -- & 1.5 & 4.25 & 11.7 & 1.1 \\
                        {\rm HM} & -- & -- & -- & 1.3 & 3 & 9 & 1.1 \\
                \end{tabular}
\caption{Material parameters used in our calculations. The energies $\mathcal{I}$ 
and $\mathcal{A}$ denote ionization and affinity levels of the indicated helium 
configurations~\cite{SM05,BTG94}, $Z_{\rm eff}$ is the effective charge used in the hydrogen-like 
wavefunctions $\psi_{1\downarrow}(\vec{r})$ and $\psi_{2\sigma}(\vec{r})$ (required for the 
calculation of the matrix elements) to reproduce these energies, $z_i$ is the position of the 
image plane, and $\phi$, $E_{\rm F}$, and $m^*_e$ are the work function, the Fermi energy, and the 
effective mass of an electron in the conduction band of the metal 
surface~\cite{HS79,LKW03,AM76,Hagstrum54a,PBF15a}.}
                \label{tableParameter}
        \end{center}
\end{table}

In this section we present numerical results calculated for the material parameters listed in 
Table~\ref{tableParameter}. We use atomic units measuring length in Bohr radii and energy in 
Hartrees. The surface is assumed to be at room temperature leading to a thermal broadening of
the Fermi-Dirac distribution which is much less than the velocity-induced smearing. In the 
calculations we used therefore \eqref{VelocitySmearing} in the limit $T_s\rightarrow 0$.

\begin{figure}[t]
        \includegraphics[width=0.9\linewidth]{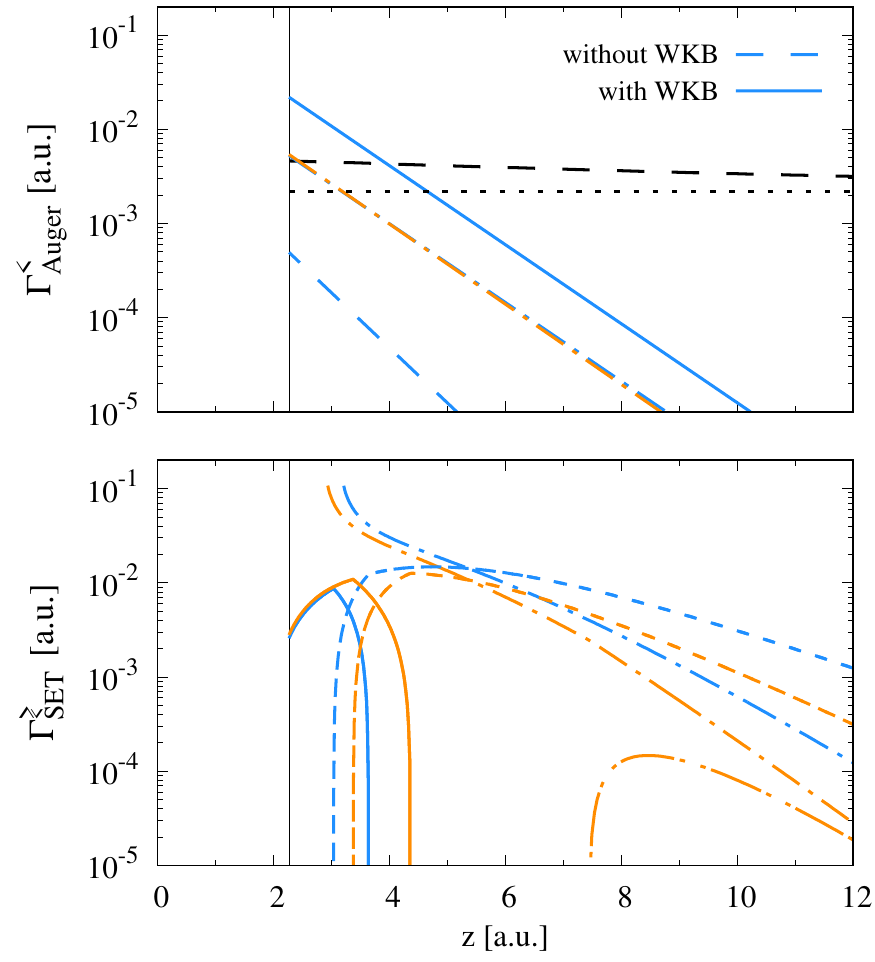}
	\caption{(Color online) Transition rates~\eqref{staticApprox2Sp}--\eqref{approxIAD} 
entering~\eqref{rateEq} for a $\HeliumPositiveIon$ ion hitting an aluminum surface with 
$E_{\rm kin}=60\,{\rm eV}$ and angle of incidence $\alpha=15^\circ$. The turning point 
$z_{\rm TP}=2.27$ is indicated by the thin vertical lines. The upper panel shows the 
rates for autodetachment $\Gamma_{\rm AuD}$ (black dotted) and the Auger processes, 
$\Gamma^<_{\rm DAD\downarrow}$ (black dashed), 
$\Gamma^<_{\rm IAD\uparrow}$ (orange dash-dotted), and $\Gamma^<_{\rm IAD\downarrow}$ 
(blue dash-dotted). The latter two turn out to be almost identical but this must 
not always be the case. In addition, $\Gamma_{\rm AN}^<$ is shown with (blue solid) 
and without (blue dashed) WKB correction. Including it 
increases $\Gamma_{\rm AN}^<$ by two orders of magnitude making it to coincide in the 
intervals most relevant for the charge transfer we discuss with the rates obtained
by other means (see Fig.~\ref{ANrateComparison} and discussion in main text). The lower panel 
presents the rates due to single-electron transfer: $\Gamma^<_\uparrow$ (orange dotted), 
$\Gamma^>_\uparrow$ (orange dash-dotted), $\Gamma^>_\downarrow$ (blue dashed-dotted), 
$\Gamma^<_{-,\uparrow}$ (orange solid), $\Gamma^<_{-,\downarrow}$ (blue solid), 
$\Gamma^>_{-,\uparrow}$ (orange dashed), and $\Gamma^>_{-,\downarrow}$ (blue dashed).
The rate $\Gamma^<_\downarrow$ is not shown. It is less than $10^{-10}$ and thus 
negligible.} 
	\label{GammaAlpic}
\end{figure}

We start the discussion with Fig.~\ref{GammaAlpic}, where we plot the transition rates 
entering the rate equation~\eqref{rateEq} for a $\HeliumPositiveIon$ ion hitting an 
aluminum surface with $E_{\rm kin}=60\,{\rm eV}$ and angle of incidence $\alpha=15^\circ$.
The upper panel shows the Auger rates whereas the rates due to single-electron 
transfer are shown in the lower panel. To demonstrate the importance of the WKB correction
to the Auger rates we plot in the upper panel $\Gamma^<_{\rm AN}(t)$ calculated with 
and without it. Clearly, the WKB correction to the metal wavefunction 
has a dramatic effect. It increases $\Gamma^<_{\rm AN}(t)$ by two orders of magnitude. 
A comparison with the results from other groups, discussed in the
next paragraph, indicates that the WKB correction is essential for 
producing the correct order of magnitude. The WKB correction is also important 
for indirect Auger de-excitation. Due to lack of data we can however not compare it 
with other results. Before discussing the reliability of the rates, a few 
general remarks are in order. The rates for indirect 
Auger de-excitation and Auger neutralization decrease with distance whereas the rate for 
direct Auger de-excitation remains almost constant. This is simply because it is a transition 
between two ionization levels which shift more or less identically. In this respect it resembles
the rate for autodetachment $\Gamma_{\rm AuD}$ which is exactly a constant within our modeling 
and moreover independent of the target surface. Comparing the Auger rates with the rates 
for single-electron transfer (plotted in the lower panel) shows that Auger rates are in general 
smaller implying that the latter dominate the former in situations where both are possible. The 
spin-dependence of the rates arises primarily from the energy difference of the singlet and 
triplet ionization/affinity levels. The closer the levels to the vacuum level the more extended 
is the wavefunction of the surface electron taking part in the process leading to a larger matrix 
element and hence transition rate. For the same reason $\Gamma^\gtrless_{-,\sigma}$ decreases 
near the turning point.

\begin{figure}[t]
        \includegraphics[width=0.95\linewidth]{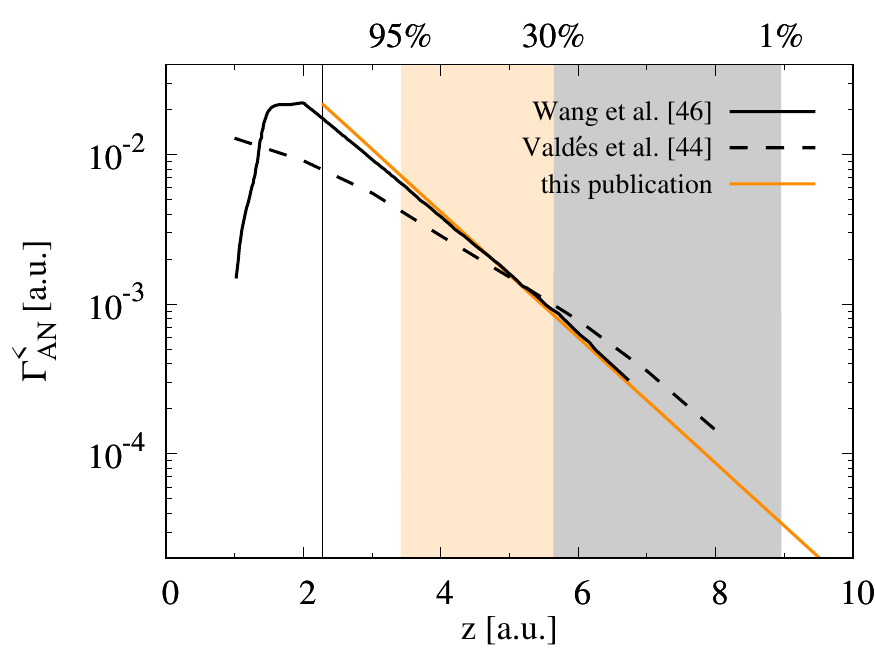}
        \caption{(Color online) Comparison of the WKB-correct Auger 
        neutralization rate with the rates obtained by Wang and coworkers~\cite{WGM01} 
        and Vald\'es et al.~\cite{VGB05}. The collision parameters are the same as in 
        Fig.~\ref{GammaAlpic}, the surface is Al(100), and the turning point 
        $z_{\rm TP}=2.27$ is indicated by the vertical line. For distances 
        $z\gtrsim 3.5$, where we shall find $95\%$ of the reaction to 
        take place for an angle of incidence of $15^\circ$, the agreement is rather good 
        although the rates have been obtained by different methods using different 
        approximations. Close to the turning point our rate (and the one of Wang and 
        coworkers) is about a factor two too large compared to the rate of Vald\'es et al. 
        which is based in part on first principles. As far as our rate is concerned we take 
        this as an indication that non-orthogonality corrections (which we neglect) are 
        already sizeable at $z \lesssim 3.5$. The percentiles of the reaction 
        change with angle of incidence. For perpendicular incidence the $95\%$ line is 
        closer to the turning point. The neglect of the corrections becomes thus more 
        important in this case. 
        }
        \label{ANrateComparison}
\end{figure}

To estimate the quality of our WKB-modified rate for Auger neutralization we compare 
it in Fig.~\ref{ANrateComparison} with the rate given by Wang and coworkers~\cite{WGM01}. 
It is essentially an extension of the Auger neutralization rate worked out by Lorente and 
Monreal~\cite{LM96} to distances $z<2\,\mathrm{a_B}$ and well established~\cite{LKW03,MMM98}.
The agreement for $z>z_{\rm TP}=2.27\,\mathrm{a_B}$ is almost perfect, although the two
rates are obtained by different methods. Additional support for our rate 
(and hence also for the one of Wang and coworkers) stems from the comparison
with the rate obtained by Vald\'es and coworkers~\cite{VGB05} using an approach based 
in part on first principles. At distances, where Auger neutralization is expected to take place, 
there is an astonishingly good agreement between the three rates indicating that the three 
approaches contain the essential physics operating at these distance. They differ hence only 
in aspects becoming important at high impact energies, when the projectile gets 
closer to the target or may even penetrate it, as can be seen by the deviations 
at short distances. Since the model assumptions are the same for the other rates we 
calculate, we except them to be also of the correct order of magnitude for $z \gtrsim 2\,\mathrm{a_B}$, 
that is, at distances where at moderate impact energies charge-transfer takes place.

\begin{figure*}[t]
%{\includegraphics[width=0.45\linewidth]{occ_W_5.pdf}}
%{\includegraphics[width=0.45\linewidth]{occ_Cu_25.pdf}} \\
%{\includegraphics[width=0.45\linewidth]{occ_Al_15.pdf}}
%{\includegraphics[width=0.45\linewidth]{occ_mix_5.pdf}}
{\includegraphics[width=0.45\linewidth]{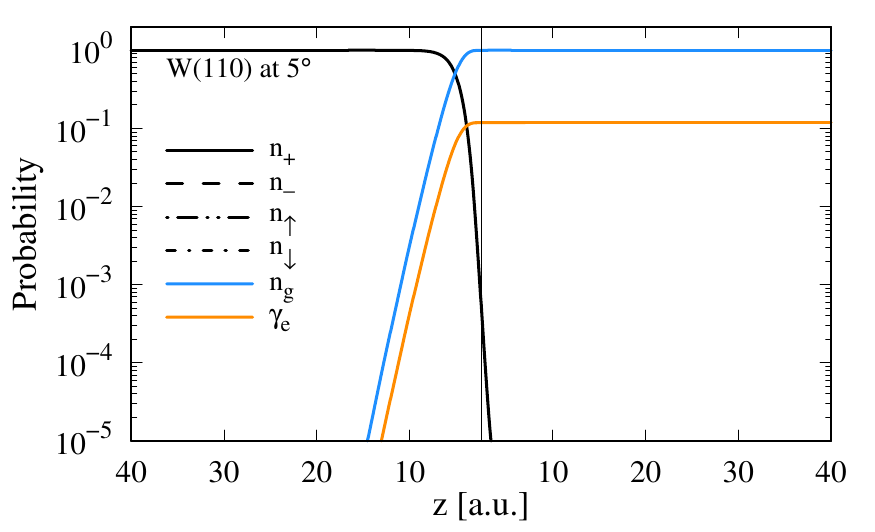}} 
{\includegraphics[width=0.45\linewidth]{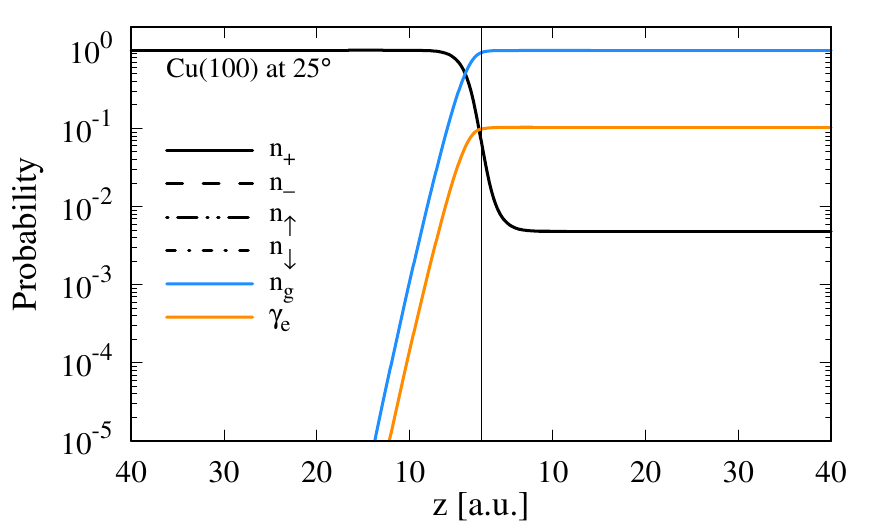}} \\
{\includegraphics[width=0.45\linewidth]{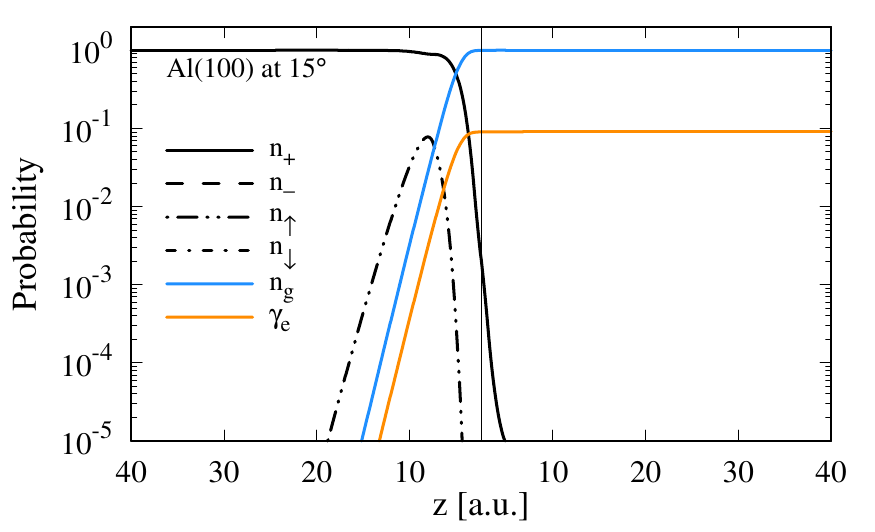}} 
{\includegraphics[width=0.45\linewidth]{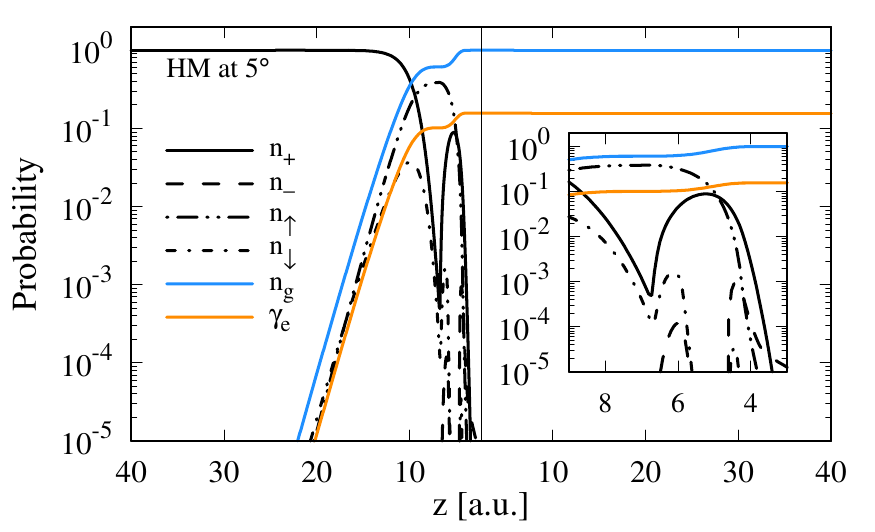}} 
\caption{(Color online) Instantaneous probabilities for electron emission and the 
occurrence of the various electronic configurations of the ${\rm He}$ projectile, 
which initially was in the $\HeliumPositiveIon$ configuration, obtained 
from~\eqref{rateEq} using the transition rates defined in Eqs. 
\eqref{staticApprox2Sp}--\eqref{staticApprox2Sm}. The species for which no data are 
shown do not affect the charge transfer. Their occurrence probabilities are less 
than $10^{-5}$ and thus negligible. The kinetic energy of the initial ${\rm He}^+$ 
ion scattering off the different surfaces is $E_{\rm kin}=50$ eV [W(110)], 
$E_{\rm kin}=25$ eV [Cu(100)], $E_{\rm kin}=60$ eV [Al(100)], and $E_{\rm kin}=50$ eV [HM]. 
The turning point $z_{\rm TP}=2.27$ is indicated by the thin vertical line 
in the middle of the plots and the inset in the lower right panel provides 
an enlarged look on the incoming branch in front of the turning point.}
\label{InstStatic}
\end{figure*}

Having calculated the transition rates we can solve the rate equation for the instantaneous occurrence
probabilities $n_+(t), n_\uparrow(t), n_\downarrow(t)$, $n_-(t)$, and $n_g(t)$, applying respectively 
to the positive ion, the triplet and singlet metastable state, the negative ion, and the groundstate. 
Figure~\ref{InstStatic} shows results for these quantities for a $\HeliumPositiveIon$ ion hitting 
different surfaces at different angles of incident and different kinetic energies. The abscissas show 
the separation of the projectile from the surface. Starting on the left at a distance $z=40$ 
it moves along the incoming branch of the trajectory towards the turning point 
$z_{\rm TP}=2.27$, indicated by the thin vertical line, where it is specularly reflected 
to move back to the distance $z=40$ along the outgoing branch of the trajectory shown on 
the right. The kinetic energy of the projectile was set to $E_{\rm kin}=50$ eV [W(110)], 
$E_{\rm kin}=25$ eV [Cu(100)], and $E_{\rm kin}=60$ eV [Al(100)] which are the kinetic 
energies at which the electron emission spectra have been determined experimentally for these 
metals~\cite{MHB93,LKW03}. Below we will compare the calculated spectra with the experimentally 
measured ones. We also studied a hypothetical metal, termed "HM", with $E_{\rm F}=9~\mathrm{eV}$ and
$\phi=3~\mathrm{eV}$ to make all processes outlined in Fig.~\ref{outlineProcesses} to work in concert
for an ion with $E_{\rm kin}=50~\mathrm{eV}$ and $\alpha=5^\circ$.
In case of tungsten and copper, the work functions, $\phi=5.22~\mathrm{eV}$ (tungsten) and  
$\phi=5.1~\mathrm{eV}$ (copper), are too large to enable resonant single-electron transfer 
into the metastable states $\HeliumMetaSing$ and $\HeliumMetaTrip$. Hence, at the end 
only the groundstate $\HeliumGroundState$ becomes occupied via Auger neutralization, with 
probability unity for tungsten and near unity for copper. The ion is thus very efficiently 
neutralized at both surfaces. For copper however the positive ion has a slim chance to survive. 
Its occurrence probability at the end of the collision $n_+(t\rightarrow\infty)\approx 0.004$. 
The secondary electron emission probability, the $\gamma-$coefficient, is for both cases 
around $0.1$. Analyzing the two cases a bit deeper one realizes that the larger angle of 
incident makes the projectile hit the copper surface with a much larger perpendicular kinetic 
energy. Since the major part of the reaction still takes place for distances $z<10$, the 
interaction time for copper is much shorter than for tungsten. This may be the reason 
for the ion to survive the collision, albeit only with a very small probability. 

For an aluminum surface, the work function is low enough to allow on the incoming branch
of the trajectory also the formation of the $\HeliumMetaTrip$ configuration. Its 
occurrence probability $n_\uparrow(t)$ raises to sizeable values around 
$z=10$ (double-dot dashed line in the lower left panel of Fig.~\ref{InstStatic}). 
Secondary electron emission due to indirect Auger de-excitation it enables is however 
very weak. We find only one percent of the total emission probability to be due to this 
process, consistent with the statement of Wang and coworkers~\cite{WGM01} that it is negligible. 
As can be seen in the lower left panel of Fig.~\ref{InstStatic}, secondary electron emission 
due to indirect Auger de-excitation becomes small compared to emission due to Auger neutralization 
because its starting point, the metastable states, are most of the time much less probable than 
the positive ion, the starting point for Auger neutralization. Hence, although the rates for 
indirect Auger de-excitation and Auger neutralization are of the same order of magnitude, 
differing only by a factor two (see Fig.~\ref{GammaAlpic}), the efficiency of the two processes 
is very different due to the collision dynamics. Iglesias--Garc\'ia and coworkers~\cite{IGG14}, 
in contrast, report on the importance of single-electron transfer, and hence the formation of 
metastable states, for the neutralization of a helium ion at an aluminum surface. The noticeable 
temporary occurrence probability we find for $\HeliumMetaTrip$ seems to support their view. However, 
its role for the outcome of the collision process is very sensitive to the position of the Fermi energy 
and the shift of the ionization level ${\cal I}_\HeliumMetaTrip$ encoded in Eq.~\eqref{2sTNshift}.
In our case, we find that at the end $\HeliumMetaTrip$ plays a subdominant role. Further investigations 
are required to clarify the issue, taking improved models for the electronic structure of the surface 
and the polarization-induced level shifts into account.

\begin{figure*}[t]
%        {\includegraphics[width=0.45\linewidth]{final_W.pdf}}
%        {\includegraphics[width=0.45\linewidth]{final_Cu.pdf}} \\
%        {\includegraphics[width=0.45\linewidth]{final_Al.pdf}}
%        {\includegraphics[width=0.45\linewidth]{final_mix.pdf}}
        {\includegraphics[width=0.45\linewidth]{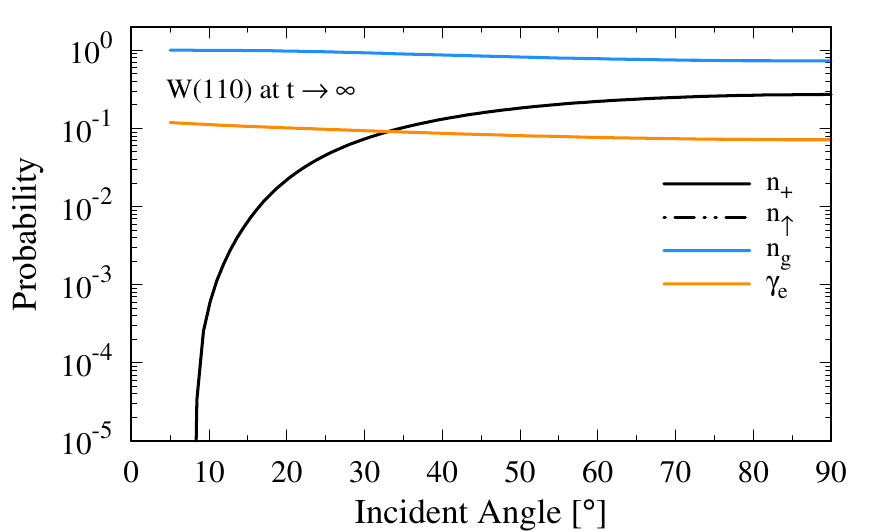}}
        {\includegraphics[width=0.45\linewidth]{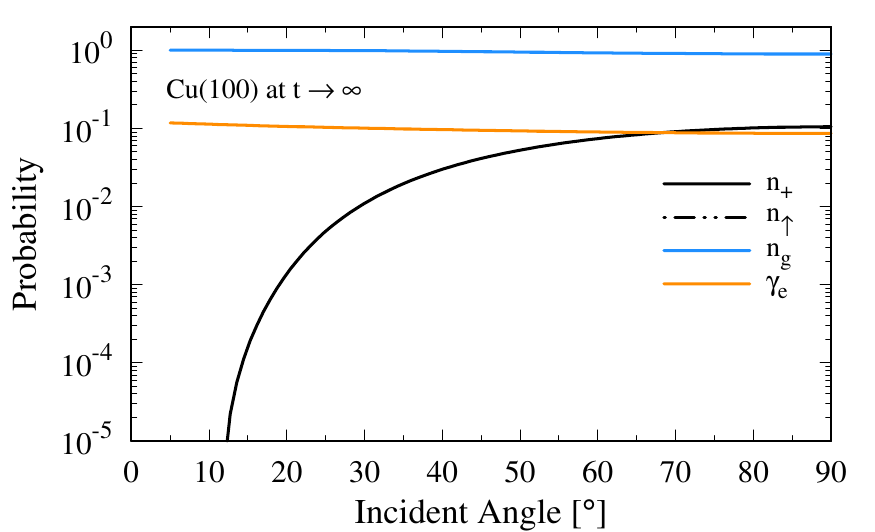}} \\
        {\includegraphics[width=0.45\linewidth]{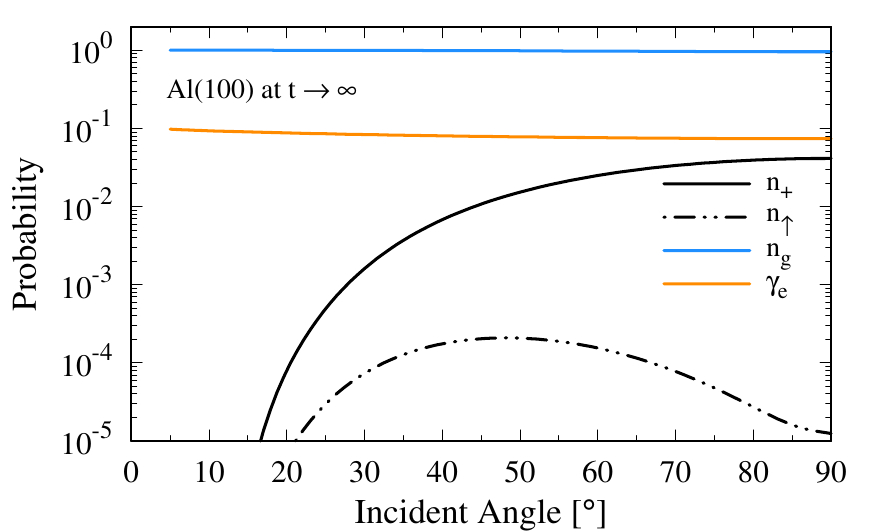}}
        {\includegraphics[width=0.45\linewidth]{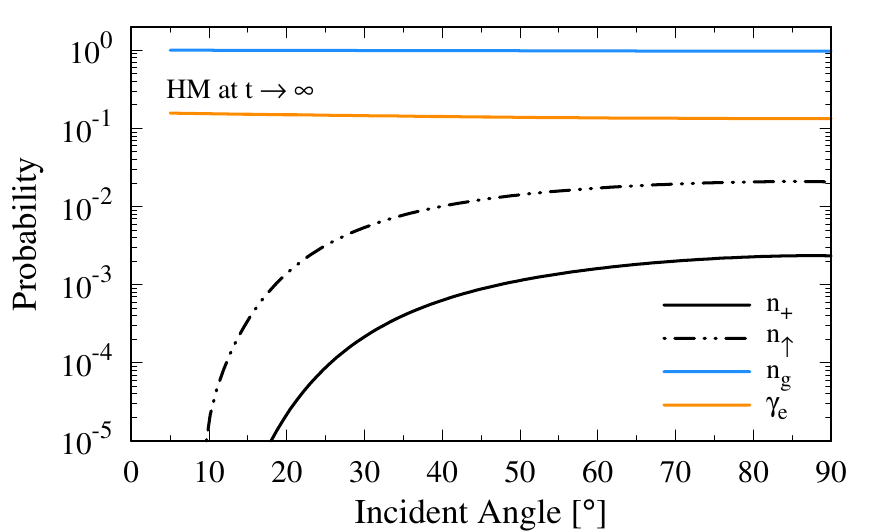}}
        \caption{(Color online) Final probabilities $\gamma_e$, $n_+$, $n_g$, and $n_{\uparrow}$ 
for electron emission and the occurrence of the $\HeliumPositiveIon$, $\HeliumGroundState$, and 
$\HeliumMetaTrip$ configurations as a function of the angle of incident. The kinetic energy of 
the helium projectile is $E_{\rm kin}=50$ eV [W(110)], $E_{\rm kin}=25$ eV [Cu(100)],
$E_{\rm kin}=60$ eV [Al(100)], and $E_{\rm kin}=50$ eV [HM]. Only the positive ion, the 
groundstate, and the triplet metastable state occur at the end of the collision with a 
noticeable probability. Negative ion and singlet metastable state are only temporarily formed. 
At the end of the collision their occurrence probabilities are vanishingly small.}
        \label{angle}
\end{figure*}

The situation we termed "HM" was constructed to demonstrate the interplay of all channels 
outlined in Fig.~\ref{outlineProcesses}. For this case, the instantaneous occupancies shown 
in the lower right panel of Fig.~\ref{InstStatic} and its inset are more involved. During 
the approach of the projectile to the surface both metastable states--$\HeliumMetaSing$ and 
$\HeliumMetaTrip$--become occupied, enabling thereby direct (from the singlet configuration) 
and indirect (from the singlet and triplet configurations) Auger de-excitation, in addition 
to Auger neutralization. The occurrence probability of the positive ion drops accordingly. 
At $z\approx 4$ before the turning point $n_+(t)$ reaches a local minimum
but starts to rise again for a brief amount of time before it drops to very small values.
At the same time the probability for $\HeliumMetaSing$ decreases after reaching its maximum. 
Having only an ionization energy of around $\mathcal{I}_\HeliumMetaSing\approx3.9~\mathrm{eV}$, 
the drop is due to the image-shift encoded in~\eqref{2sSshift} which pushes the ionization 
level above the Fermi energy thereby turning the weak gain due to single-electron transfer 
off and the strong electron loss due to the process on. In addition, there is a strong loss 
due to direct Auger de-excitation. The triplet configuration $\HeliumMetaTrip$ is affected 
similarly, albeit at a later time due to the greater ionization energy and the lacking of the 
strong direct Auger de-excitation (which is absent because of the Pauli principle). When the electron 
transfers from the metastable states back to the surface via single-electron transfer the positive 
ion is restored. Hence, the occurrence probability for the positive ion rises again near the surface, 
allowing for a revival of the Auger neutralization. As a result, the occurrence probability $n_g(t)$ 
jumps close to the surface to near unity. With the ionization levels shift of course also the affinity 
levels. If they approach the Fermi energy from above, a negative ion becomes possible. Hence, for a very 
short time interval, when the occurrence probabilities for the two metastable configurations are 
already decreasing, a negative ion is formed. It decays however nearly instantly because of 
single-electron transfer and autodetachment.
 
In all four cases depicted in Fig.~\ref{InstStatic}, the outgoing branch lacks complex behavior. For 
the chosen angles of incident and kinetic energies the groundstate is always formed very efficiently 
along the incoming branch. Since the groundstate is not subject to a loss channel, it cannot be 
destroyed. The constraint~\eqref{constraint}, which has to be satisfied at any instant of time, 
ensures then that the other configurations vanish as soon as the groundstate appears with probability 
near unity. At the end of the collision the groundstate configuration dominates. Only for copper 
we find a noticeable probability for detecting at the end also a positive ion. Although the other 
configurations have vanishingly small probabilities at the end they may nevertheless affect the 
outcome of the collision because of their presence at intermediate times.

Experimentally accessible are only the probabilities at the end of the collision. Let us 
thus investigate their dependence on impact energy and angle of incidence. Figure~\ref{angle} 
shows for the same impact energies as in Fig.~\ref{InstStatic} the angle dependence of the 
probabilities for detecting at the end of the collision the configurations included into our 
modeling as well as for emitting an electron. The results for tungsten and copper are again very similar. 
At small angles essentially only the groundstate is formed, because Auger neutralization is the dominant 
process. As the angle increases the kinetic energy perpendicular to the surface also increases, lowering 
thereby for all channels the interaction time. This leads to a steady increase of the occurrence 
probability for the positive ion although it remains for all angles much smaller than the probability 
for the groundstate. The ion survival probability is largest for perpendicular 
incidence, which is also most relevant for plasma applications. For tungsten we obtain 
around $0.3$, which is two orders of magnitude too large compared to the experimental data
Hagstrum~\cite{Hagstrum61} found long time ago. But survival probabilities on the order of 
$10^{-3}$ are typical (see for, instance, Fig.~26 in Ref.~\cite{Monreal14}).
Moving the turning point closer to the surface reduces the survival probability but not 
by two orders of magnitude. It is not possible to push this number to the correct order of magnitude 
by simply adjusting model parameters. We expect the neglect of single-electron transfer to the 
$\mathrm{1s}$ shell to be responsible for the too large survival probability at perpendicular incidence. 
The impact energy of the $\HeliumPositiveIon$ projectile is in this case the highest leading to the 
closest encounter with the surface where single-electron transfer from core levels may already 
become important. To include it is however beyond the scope of the present work. In addition 
non-orthogonality corrections to the Auger rates may become also an issue for perpendicular 
incidence. 

The final probabilities for aluminum and the hypothetical metal, shown in the lower two panels, 
behave also similarly. The main difference to tungsten and copper is the formation of the metastable 
triplet state $\HeliumMetaTrip$. It forms on the incoming branch of the trajectory because the 
lowering of the work function enables single-electron transfer into the metastable state and the 
shortening of the interaction time reduces the electron transfer back to the metal which, in effect,
leads to a freezing-in of the metastable state. For the hypothetical case the occurrence probability 
for the metastable triplet state is even larger than the one for the positive ion indicating that at 
intermediate times the singlet metastable state as well as the negative ion state may have also played 
an active role in the collision. 
 
\begin{figure*}[t]
%{\includegraphics[width=0.45\linewidth]{spec_W_5.pdf}}
%{\includegraphics[width=0.45\linewidth]{spec_Cu_25.pdf}} \\
%{\includegraphics[width=0.45\linewidth]{spec_Al_15.pdf}}
%{\includegraphics[width=0.45\linewidth]{spec_mix_5.pdf}}
{\includegraphics[width=0.45\linewidth]{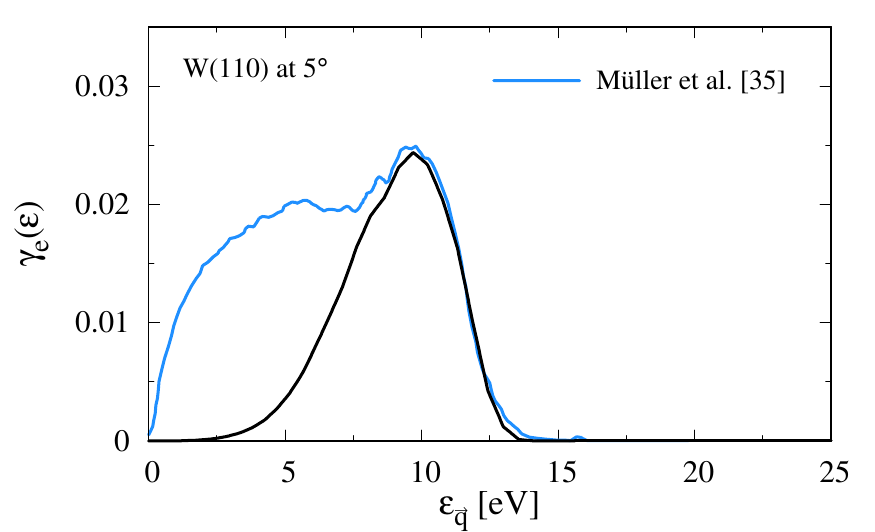}} 
{\includegraphics[width=0.45\linewidth]{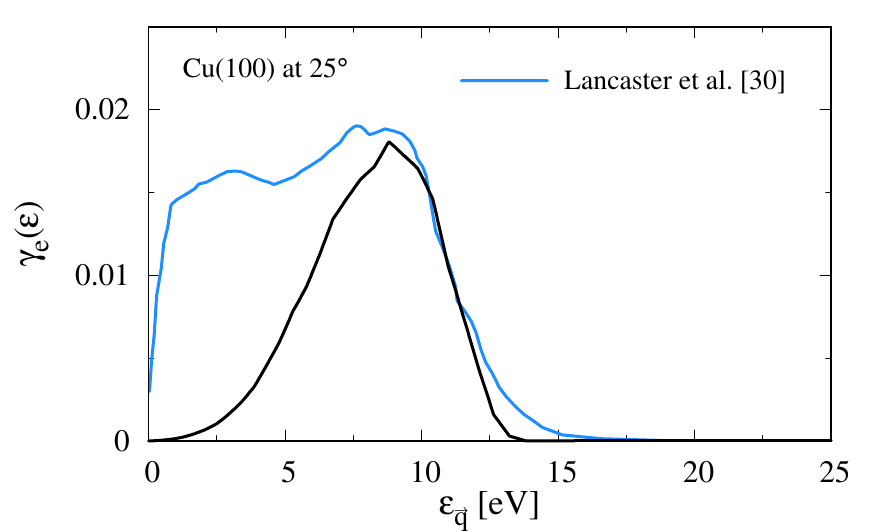}} \\
{\includegraphics[width=0.45\linewidth]{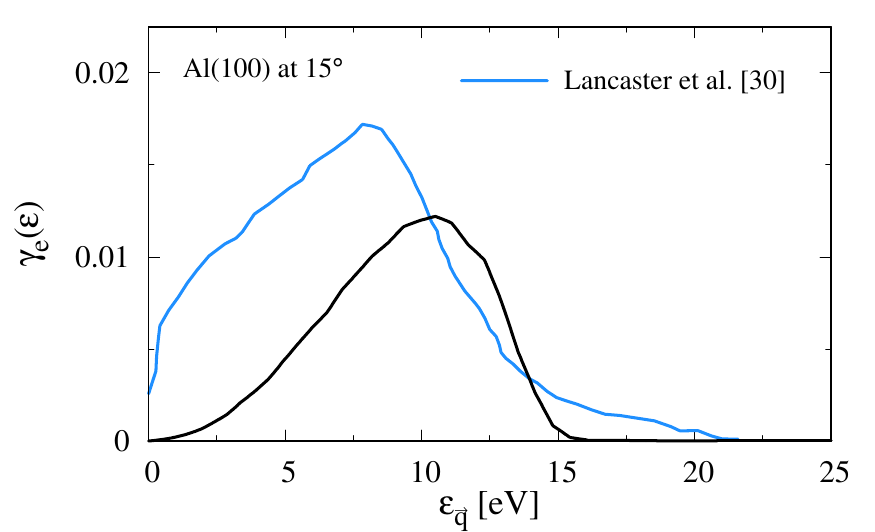}} 
{\includegraphics[width=0.45\linewidth]{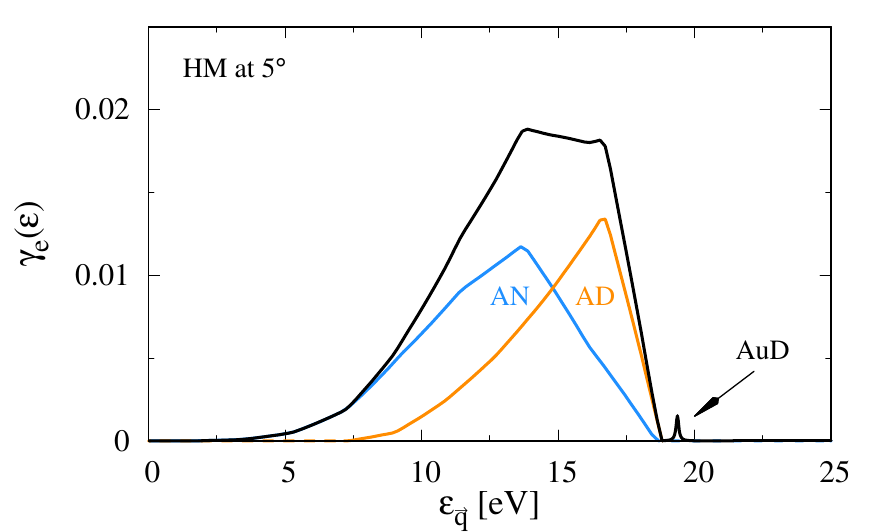}} 
\caption{(Color online) Energy spectrum of the emitted electron once the 
collision is completed. The kinetic energy of the initial ${\rm He}^+$ ion is 
$E_{\rm kin}=50$ eV [W(110)], $E_{\rm kin}=25$ eV [Cu(100)], $E_{\rm kin}=60$ eV [Al(100)],
and $E_{\rm kin}=50$ eV [HM]. The experimental data for tungsten~\cite{MHB93} 
are weighted to the electron emission coefficient $\gamma^{\rm exp}_e=0.22$ found 
experimentally. For the aluminum and copper data~\cite{LKW03} this was not possible because 
no estimates are given for the $\gamma$-coefficients. The weighting of the experimental 
data for copper and aluminum was thus performed using the tungsten data as described in 
the main text. We then obtain $\gamma^{\rm exp}_e\approx 0.19$ and 
$\gamma^{\rm exp}_e\approx 0.18$ for copper and aluminum, respectively.} 
	\label{spectralDensity}
\end{figure*}
 
We now turn to the energy spectrum of the emitted electron. In Fig.~\ref{spectralDensity} we 
present results based on Eqs.~\eqref{secE} and \eqref{specDensity} together with experimental 
data for tungsten from M\"uller and coworkers~\cite{MHB93} and copper and aluminum from 
Lancaster and coworkers~\cite{LKW03}. Only the former group gives also an estimate for 
the total emission probability, that is, the $\gamma-$coefficient. As far as the data for 
tungsten are concerned we can thus compare absolute numbers. For copper and aluminum this 
is not possible since no value for the $\gamma-$coefficient was given by the 
experimentalists. In addition, the area embraced by the measured emission spectra, which 
would give the emission coefficient according to~\eqref{numSecE}, cannot be used either 
because the experimental data are presented in arbitrary units. 

M\"uller and coworkers estimate $\gamma^{\rm exp}_e = 0.22$ for a $\HeliumPositiveIon$ 
ion hitting a tungsten surface with $E_{\rm kin}=50$ eV and $\alpha=5^\circ$. We 
weighted their emission spectrum according to~\eqref{numSecE} to match this number. A 
comparison of the weighted experimental spectrum with our data is shown in 
Fig.~\ref{spectralDensity}. The agreement is quite satisfying, in particular, as far
as the high-energy side of the spectrum is concerned. The high-energy cut-off and 
the maximum of the emission spectrum match quite well indicating that our approach 
may be able to estimate at least the order of magnitude of secondary electron emission
in cases where no experimental data are available. 
At low energies experimental and theoretical data deviate. The theoretical secondary 
electron emission coefficient $\gamma^{\rm theo}_e=0.12$ is thus roughly only one-half
of the experimental estimate. The reason is the following: We did not include processes 
relaxing the energy of the excited electron. Scattering cascades~\cite{Propst63,LKW07}
and higher order Auger processes~\cite{BMN92} involving more than two electrons are 
often attributed for this. Since the physical origin is not yet quite clear, we did not 
consider it for the purpose of this work. Our energy spectrum for the secondary electron 
leaving the tungsten surface is entirely due to Auger neutralization; the other channels 
of Fig.\ref{outlineProcesses} are energetically closed. Without scattering cascades 
and higher-order Auger processes included, it applies only to the high-energy side of 
the spectrum. There, however, the agreement is rather good.

The good match of the theoretical and experimental emission spectra for tungsten at high 
energies suggests a way to scale the data of Lancaster and coworkers such that 
they can be compared to the calculated spectra. An important consequence of 
the scaling is that we can then also estimate the $\gamma-$coefficients for copper and 
aluminum. The ratio $r=\gamma_e^{\rm theo}/\gamma_e^{\rm exp}$ of the theoretical and 
experimental secondary electron emission coefficients for tungsten is roughly one-half 
because of the neglect of scattering cascades and higher-order Auger processes. Assuming 
that both types of processes are essentially the same for the metals under discussion, 
we scale the emission spectra of Lancaster and coworker also in this manner. Hence, we 
set $\int dE \gamma_e(E,t\rightarrow \infty)/\int dE \gamma_e^{\rm exp}(E)=r$ 
where $r$ is the ratio obtained from the tungsten data. The scaling provides an absolute
scale to the experimental data and hence also the $\gamma$-coefficients.

As can be seen from Fig.~\ref{spectralDensity}, applying the scaling to the data for copper 
leads at high energies again to a good agreement between the experimental and the theoretical 
emission spectra. As for tungsten, the high-energy side of the spectrum is determined largely 
by Auger neutralization. From the calculation we obtain for copper $\gamma^{\rm theo}_e=0.1$
producing $\gamma^{\rm exp}_e\approx 0.19$. For aluminum the matching of the high energy tails 
is not as good. The small work function and the large Fermi energy lead in this case to a 
broad spectrum for the electron emitted by Auger neutralization. In addition, the low work 
function enables indirect Auger de-excitation although it provides only a small amount of 
secondary electrons between $15$ and $20\,\mathrm{eV}$. For aluminum our approach yields 
$\gamma^{\rm theo}_e=0.09$, a bit lower than for tungsten and copper. The estimate for 
the experimental value is thus $\gamma^{\rm exp}_e\approx 0.18$. For aluminum the 
theoretically obtained emission spectrum does not even match the measured data at high 
energies. In the case of tungsten and copper the processes leading to electron emission at lower 
energies are well separated from electron emission due to Auger neutralization. The latter 
leading to a maximum at the high-energy side while the former producing a flat low-energy 
shoulder. The experimental data for aluminum in contrast feature a single asymmetric emission 
peak suggesting that Auger neutralization and the low-energy processes strongly overlap. It is 
thus clear that the scaling deduced from the tungsten data necessarily produces for aluminum 
a maximum in the experimental data which is above the maximum of the calculated spectrum. 
In order to achieve better agreement between theory and experiment the modeling has thus
to include also the processes leading to electron emission at low energies. This is however 
beyond the scope of the present work.

The analysis of the experimental emission spectra indicates that Auger neutralization 
is by far the most important process listed in Fig.~\ref{outlineProcesses}. In the 
spectra we find no features which could be attributed to Auger de-excitation or 
autodetachment. To demonstrate how these processes may in principle affect secondary 
electron emission, we constructed therefore the hypothetical metal termed "HM". Its 
secondary electron emission spectrum is shown in the lower right panel of
Fig.~\ref{spectralDensity}. With the processes of Fig.~\ref{outlineProcesses} 
simultaneously active the emission spectrum becomes asymmetric. Decomposing the 
spectrum into the contributions originating from Auger neutralization, direct and 
in-direct Auger de-excitation, and autodetachment shows that Auger de-excitation is 
responsible for the steep high-energy cut-off whereas Auger neutralization gives 
rise to the low-energy tail of the emission spectrum. Autodetachment adds only 
a faint peak above the main feature. Our model is however not able to get the 
autodetachment peak at the energy expected from other studies~\cite{HC91,BTG94}. 
Most probably this is due to the incompleteness of the level shifts. In 
addition to the shifts induced by the image interaction there are contributions 
arising from the non-orthogonality of the surface and projectile states. To include 
them was however also beyond the scope of the present work.

\section{Conclusions}
\label{Conclusions}

In this work we presented a generic quantum-kinetic approach for calculating the 
probability with which a secondary electron arises due to the neutralization of
a positive ion on a surface as well as the energy with which it emerges. Focusing 
on impact energies where the internal potential energy of the projectile drives 
the emission and taking a $\HeliumPositiveIon$ ion hitting a metal surface as an 
example we showed that the approach is capable to treat the three main emission 
channels on an equal footing which may be open in this energy range: Auger 
neutralization to the projectile's groundstate, single-electron transfers to 
excited (metastable) states followed either by indirect/direct Auger de-excitation, 
in case the states are neutral, or autodetachment in case the states are negatively 
charged. 

The approach is based on a semiempirical Anderson-Newns model. It describes the 
projectile by a time-dependent few-level system and the target surface by a step 
potential. Parameterizing the few-level system and the step potential by experimental 
values for the energy levels involved (work function, Fermi energy, electron affinities, 
and ionization energies) and employing models for energy shifts and approximate 
wavefunctions of the correct symmetry for the calculation of matrix elements it provides 
a flexible tool for describing charge-transferring atom-surface collisions. It can be 
applied to any projectile-target combination. In particular, it is not restricted to 
ideal surfaces or to a particular crystallographic orientation of the surface. Both can 
be taken into account by a suitable choice of the work function and the Fermi energy. To 
implement an Anderson-Newns model for a charge-transferring atom-surface collision it 
is thus necessary (i) to identify the ionization and affinity levels which may become 
active in the charge transfer, (ii) to parameterize and furnish the model as described 
above, and (iii) to calculate the matrix elements. After the model is constructed the 
analysis of the charge-transfer proceeds in a canonical manner using the quantum-kinetic 
framework of contour-ordered Green functions. An advantage of the approach is thus that 
it separates the quantum kinetics of charge-transfer from the many-body theoretical 
description of the non-interacting projectile and target. The latter is simply encoded 
in the matrix elements of the model Hamiltonian serving as the starting point to the 
former. Had the matrix elements been obtained by a different method--for instance, by 
an ab-initio density functional approach--the quantum kinetics would be the same.

To model the helium projectile we constructed an effective three-level system. 
It represents the groundstate $\HeliumGroundState$, the singlet and triplet 
metastable states, $\HeliumMetaSing$ and $\HeliumMetaTrip$, and the negative
ion $\HeliumNegativeIon$. The ionization and affinity levels associated with 
these states shift while the projectile approaches and retreats from the surface. 
The energies of the three levels are thus time-dependent. We mimic these 
dependencies by polarization-induced image shifts. At short distances corrections
to the shifts occur due to the non-orthogonality of the surface and target 
wavefunctions. Since in the situations we have studied the charge transfer occurs 
preferentially at relatively large distances from the surface we did not include 
the corrections in the present work. Discrepancies we found between calculated and 
measured data indicate however that they have to be included in the future. 
The matrix elements coupling the projectile and the target depend also 
on time. To obtain numerical values for them we approximated the electron wavefunctions 
of the surface by the wavefunctions of the step potential and the electron wavefunctions 
of the helium projectile by screened $1s$ and $2s$ hydrogen wavefunctions. A comparison
with helium and lithium Roothaan-Hartree-Fock wavefunctions indicated that this approximation,
which enables at least in part an analytical treatment of the matrix elements, is 
justified. In fact, it turns out that the rate for Auger neutralization 
we obtain from the hydrogen-like wavefunction for the projectile's $\mathrm{1s}$ 
shell and the wave functions of the step potential is in good agreement with the 
rate obtained by an investigation based at least in part on ab-initio methods, if the
tunneling of the surface electron filling the hole in the $\mathrm{1s}$ shell is taken 
into account semiclassically by a WKB correction. We included the corrections in the 
other Auger matrix elements as well where tunneling through the barrier takes place. We 
expect them to be thus also of the correct order of magnitude. 

Essential for an efficient handling of the few-level system is the use 
of projection operators and auxiliary boson(s). The projection operators 
allow to account within the same few-level system for projectile states with
different internal energies--assigning different energies to the levels 
depending on the occupancy and hence the configuration represented--while 
the auxiliary boson(s) allow to switch without violating energy conservation
between the configurations as required by the interactions included in the 
Hamiltonian. For the helium projectile two auxiliary bosons are needed. 
Other projectiles may require more than two. Applying the technique to the 
helium projectile enabled us to treat Auger neutralization, Auger de-excitation,
and auto-detachment on an equal footing. For the quantum kinetic analysis
of the collision dynamics the projection operators are rewritten in terms of 
pseudo-particle operators. It is then straightforward--using standard techniques 
of many-body theory--to set up the quantum kinetic approach from which the 
rate equation is obtained for the probabilities with which the projectile 
configurations occur and an electron is emitted in the course of the collision.

The rate equation follows from a saddle-point approximation to the equations
of motions for the occurrence probabilities in the non-crossing approximation,
which is sufficient because we do not expect Kondo-type correlations to occur 
on the projectile under typical plasma conditions. In addition to these 
approximations we postulated an approximate factorization of 
the $t$ and $\vec{k}$-dependence of the matrix elements to stabilize and speed-up 
the numerics. At the moment, due to the absence of exact expressions for the 
transition rates, its validity cannot be verified. However, the final results 
for the occurrence probabilities and the secondary electron emission coefficient
compare favorable with measured data, with differences attributable to 
physical processes not included in the modeling, suggesting that the factorization
is not too critical.

The numerical solution of the rate equations showed that the occurrence probabilities 
for the projectile configurations are determined along the incoming branch of the 
collision trajectory. On the outgoing branch they essentially do not change anymore. 
This is the case because there is no channel leading from the groundstate 
$\HeliumGroundState$ back to any of the other configurations considered in the 
model. The angle dependencies of the final occurrence and electron emission 
probabilities show that for perpendicular incident--the case most relevant for 
plasma walls--the projectile has a small chance for returning as a positive ion 
after having induced a secondary electron. Due to the neglect 
of single-electron transfer from deeper lying levels of the surface to the 
$\mathrm{1s}$ shell we found the ion survival probability however two orders of magnitude 
too large compared to experimental values. The $\gamma$-coefficient we obtain is much 
better. It is only a factor two too small compared to experimental data (where available). 
The discrepancy arises from the neglect of higher-order Auger processes and/or 
scattering cascades. That these processes 
are important we deduced from an analysis of the emission spectra. At high 
energies the spectra we obtained compare favorably with experimental data from 
different groups, especially for tungsten. The mismatch is at low energies 
where one expects higher-order Auger processes and/or inelastic scattering 
cascades to affect the emission spectra. Using the ratio of the calculated and 
measured secondary electron emission coefficients for tungsten, we tried to 
quantify the contribution of the neglected processes to the secondary electron 
emission coefficient. We found that they roughly lead to its doubling. The ratio
we also used to scale the measured emission spectra for copper and aluminum,
which were given in arbitrary units. As a result we could estimate the secondary 
electron emission coefficient also for these two metals. The 
results obtained are quite reasonable indicating that our approach may have the potential 
not only for qualitative studies of ion-induced secondary electron emission but also 
for producing quantitative data, giving at least estimates of the correct 
order of magnitude. Although we included all three possible emission channels, 
for the metals investigated Auger neutralization turned out to be always the 
dominant one for electron emission. The work functions being simply too large 
for an efficient direct/indirect Auger de-excitation to take place.

For plasma applications a compact formula for the secondary electron emission 
coefficient would be very useful. Due to the complexity of charge-transferring 
atom-surface collisions and their non-universality it is however unlikely
to exist. What could be hoped for instead is a semiempirical description of 
the charge-transfer processes, adjustable to various situations of interest.
Based on the results presented in this work we identify four main issues which 
have to be tackled in order to achieve such a description: (i) Non-orthogonality 
corrections to the level shifts and Auger rates at short projectile-target separations 
should be included. This is particularly important for 
processes involving metastable configurations of the projectile. (ii) Single-electron
transfer from deeper lying states of the surface to the projectile's groundstate 
should be taken into account. This is important for obtaining realistic values for 
the ion survival probability. (iii) Energy loss of the escaping electron due to scattering 
cascades and higher-order Auger processes should be considered in order to obtain the 
energy spectrum of the emitted electron also correct at low energies. 
(iv) Approximation schemes should be developed for the high-dimensional integrals 
defining the transition rates in terms of the matrix elements which are accurate 
and numerically efficient.

Additional issues, which we consider however less critical 
because they can be overcome on the expense of additional numerical burden, without
changing the organization of the calculation, are the use of (effective) hydrogen 
wavefunctions for the projectile and the potential step for the surface potential. 
The former can be replaced by other wavefunctions of quantum-chemistry, if not available 
for the considered projectile they have to be worked out, while the latter can be replaced 
by another potential which most probably implies however a numerical construction of the
surface wavefunctions. Our results suggest however that the gain due to these 
modifications is most probably small. One has to address the four main issues to 
make a significant step forward.

Not all materials presently used as 
plasma walls require to include simultaneously Auger neutralization, Auger de-excitation, 
and auto-detachment. But having a formalism capable of doing it will enable one 
to explore the possibility of engineering the spectrum of the emitted electron 
by judiciously modifying the surface opening-up or closing-down thereby one or 
the other channel. With this goal in mind we developed the multi-channel approach 
for calculating secondary electron emission coefficients and spectra described in this 
work.

\section*{Acknowledgements}
M. P. was funded by the federal state of Mecklenburg-Western Pomerania through a postgraduate 
scholarship within the International Helmholtz Graduate School for Plasma Physics. In addition, 
support from the Deutsche Forschungsgemeinschaft through project B10 of the Transregional 
Collaborative Research Center SFB/TRR24 is greatly acknowledged.\\

\appendix 

\section{Selfenergies\label{AppSelf}}

In this appendix the selfenergies are listed from which we calculate the   
rates~\eqref{staticApprox2Sp}--\eqref{approxAuD}. The 
selfenergies, implicitly defined in~\eqref{rateEqEa}--\eqref{rateEqDa}, are 
all constructed in the non-crossing approximation as suggested by Langreth
and coworkers~\cite{LN91,SLN94a,SLN94b}. It contains the leading contributions 
of the second-order (in the interaction matrix elements) selfenergies renormalizing 
the Green functions of the projectile. First order corrections are absent. Vertex
renormalizations (diagrams with crossed lines) are ignored but they are relevant only 
in situations where Kondo-type correlations~\cite{Hewson93} occur. For surfaces 
in contact with a plasma we do not expect this.

The fermionic selfenergies $\Pi_{e,d}(t,t^\prime)$, belonging to the fermionic propagators 
$E(t,t^\prime)$ and $D(t,t^\prime)$, respectively, apply to the empty and doubly filled 
configurations, that is, the positive and negative ion. Likewise the bosonic selfenergies
$\Sigma_{n\sigma}(t,t^\prime)$, belonging to the bosonic propagators $S_{n\sigma}(t,t^\prime)$, 
apply to the configurations with a single electron, that is, the 
groundstate and the metastable states. In the expressions to follow, 
the sums in the various terms indicate their physical origin: Sums over
$\vec{k}$ are contributions due to single-electron transfer, sums over 
$\vec{k}\vec{k}'$ due to direct Auger de-excitation, sums over $\vec{k}\vec{q}$ 
due to indirect Auger de-excitation, sums over $\vec{q}$ due to autodetachment, 
and sums over $\vec{k}_1\vec{k}_2\vec{k}'$ due to Auger neutralization. Using  
Langreth-Wilkins rules~\cite{LW72}, the greater-than and less-than selfenergies obtained 
from diagrams of the type shown in Fig.~\ref{diagram} are
\begin{widetext}
\begin{align}
\label{SEglE}
	\Pi^\gtrless_e (t,t^\prime) &= \sum_{\vec{k}_1\vec{k}_2\vec{k}^\prime \sigma} V^\ast_{\vec{k}_1\vec{k}_2\vec{k}^\prime \sigma}(t) V_{\vec{k}_1\vec{k}_2\vec{k}^\prime \sigma}(t^\prime) S_{1\downarrow}^\gtrless(t,t^\prime) G^\lessgtr_{\vec{k}_1\downarrow}(t^\prime,t) G^\lessgtr_{\vec{k}_2\sigma}(t^\prime,t) G^\gtrless_{\vec{k}^\prime\sigma}(t,t^\prime) + \sum_{\vec{k}\sigma} V_{\vec{k}\sigma}(t) V^\ast_{\vec{k}\sigma}(t^\prime) S^\gtrless_{2\sigma} (t,t^\prime) G^\lessgtr_{\vec{k}\sigma}(t^\prime,t)~,\\
\label{SEglD}
	\Pi^\gtrless_d(t,t^\prime) &= \sum_{\vec{k}\sigma} V^\ast_{\vec{k}\sigma}(t) 	V_{\vec{k}\sigma}(t^\prime) S^\gtrless_{2-\sigma}(t,t^\prime) B^\gtrless_\sigma(t,t^\prime) G^\gtrless_{\vec{k}\sigma}(t,t^\prime) + \sum_{\vec{q}} V^\ast_{\vec{q}} V_{\vec{q}}\, S^\gtrless_{1\downarrow} (t,t^\prime) G^\gtrless_{\vec{q}\uparrow}(t,t^\prime)~,\\
\label{SEglS1}
	\Sigma^\gtrless_{1\downarrow}(t,t^\prime) &= \sum_{\vec{k}_1\vec{k}_2\vec{k}^\prime\sigma} V_{\vec{k}_1\vec{k}_2\vec{k}^\prime\sigma}(t) V_{\vec{k}_1\vec{k}_2\vec{k}^\prime\sigma}^\ast(t^\prime) E^\gtrless (t,t^\prime) G^\gtrless_{\vec{k}_1\downarrow}(t,t^\prime) G^\gtrless_{\vec{k}_2\sigma}(t,t^\prime) G^\lessgtr_{\vec{k}^\prime\sigma}(t^\prime,t) + \sum_{\vec{q}} V_{\vec{q}} V^\ast_{\vec{q}} \, D^\gtrless(t,t^\prime) G^\lessgtr_{\vec{q}\uparrow}(t^\prime,t) \nonumber\\
	&+ \sum_{\vec{k}\vec{k}^\prime\sigma} V_{\vec{k}\vec{k}^\prime\sigma}(t) V^\ast_{\vec{k}\vec{k}^\prime\sigma}(t^\prime) S^\gtrless_{2\downarrow}(t,t^\prime) G^\gtrless_{\vec{k}\sigma}(t,t^\prime) G^\lessgtr_{\vec{k}^\prime\sigma}(t^\prime,t) + \sum_{\vec{k}\vec{q}\sigma} V_{\vec{k}\vec{q}\sigma}(t) V^\ast_{\vec{k}\vec{q}\sigma}(t^\prime) S^\gtrless_{2\sigma}(t,t^\prime) G^\gtrless_{\vec{k}\downarrow}(t,t^\prime) G^\lessgtr_{\vec{q}\sigma}(t^\prime,t)~,\\
\label{SEglS2}	
	\Sigma^\gtrless_{2\sigma}(t,t^\prime) &= \sum_{\vec{k}} V^\ast_{\vec{k}\sigma}(t) V_{\vec{k}\sigma}(t^\prime) E^\gtrless(t,t^\prime) G^\gtrless_{\vec{k}\sigma}(t,t^\prime) + \sum_{\vec{k}} V_{\vec{k}-\sigma}(t) V^\ast_{\vec{k}-\sigma}(t^\prime) D^\gtrless(t,t^\prime) B^\lessgtr_{-\sigma}(t^\prime,t) G^\lessgtr_{\vec{k}-\sigma}(t^\prime,t)  \nonumber\\
	&+ \delta_{\downarrow \sigma} \sum_{\vec{k}\vec{k}^\prime\sigma^\prime} V^\ast_{\vec{k}\vec{k}^\prime\sigma^\prime}(t) V_{\vec{k}\vec{k}^\prime\sigma^\prime}(t^\prime) S^\gtrless_{1\downarrow}(t,t^\prime) G^\lessgtr_{\vec{k}\sigma^\prime}(t^\prime,t) G^\gtrless_{\vec{k}^\prime\sigma^\prime}(t,t^\prime) + \sum_{\vec{k}\vec{q}} V^\ast_{\vec{k}\vec{q}\sigma}(t) V_{\vec{k}\vec{q}\sigma}(t^\prime) S^\gtrless_{1\downarrow}(t,t^\prime) G^\lessgtr_{\vec{k}\downarrow}(t^\prime,t) G^\gtrless_{\vec{q}\sigma}(t,t^\prime)~.
\end{align}
\end{widetext}
The retarded selfenergies, which we also need, can be obtained from these expressions 
using the identity
\begin{align}
i H^R (t,t') = \theta (t-t') \big[ H^>(t,t') \pm H^<(t,t') \big] \,,
\label{Hfct}
\end{align}
where the minus sign applies to bosons. As explained by Langreth and 
coworkers~\cite{LN91,SLN94a,SLN94b} by applying~\eqref{Hfct} to the selfenergies 
listed above it has to be kept in mind that retarded Green functions are of order 
$Q^0$ while less-than and greater-than Green functions are of order $Q^1$ where 
$Q$ defined in~\eqref{completeness} is the projector accounting for the completeness 
of the projectile states. Hence, by constructing the retarded selfenergies via \eqref{Hfct} 
only terms should be kept which at the end lead to contributions $\propto Q^0$.
The result is 
\begin{widetext}
\begin{align}
\label{SErE}
	\Pi^R_e (t,t^\prime) &= \sum_{\vec{k}_1\vec{k}_2\vec{k}^\prime \sigma} V^\ast_{\vec{k}_1\vec{k}_2\vec{k}^\prime \sigma}(t) V_{\vec{k}_1\vec{k}_2\vec{k}^\prime \sigma}(t^\prime) S_{1\downarrow}^R(t,t^\prime) G^<_{\vec{k}_1\downarrow}(t^\prime,t) G^<_{\vec{k}_2\sigma}(t^\prime,t) G^>_{\vec{k}^\prime\sigma}(t,t^\prime) + \sum_{\vec{k}\sigma} V_{\vec{k}\sigma}(t) V^\ast_{\vec{k}\sigma}(t^\prime) S^R_{2\sigma} (t,t^\prime) G^<_{\vec{k}\sigma}(t^\prime,t)~,\\
\label{SErD}
	\Pi^R_d(t,t^\prime) &= \sum_{\vec{k}\sigma} V^\ast_{\vec{k}\sigma}(t) V_{\vec{k}\sigma}(t^\prime) S^R_{2-\sigma}(t,t^\prime) B^>_\sigma(t,t^\prime) G^>_{\vec{k}\sigma}(t,t^\prime) + \sum_{\vec{q}} V^\ast_{\vec{q}} V_{\vec{q}}\,S^R_{1\downarrow} (t,t^\prime) G^>_{\vec{q}\uparrow}(t,t^\prime)~,\\
\label{SErS1}
	\Sigma^R_{1\downarrow}(t,t^\prime) &= \sum_{\vec{k}_1\vec{k}_2\vec{k}^\prime\sigma} V_{\vec{k}_1\vec{k}_2\vec{k}^\prime\sigma}(t) V_{\vec{k}_1\vec{k}_2\vec{k}^\prime\sigma}^\ast(t^\prime) E^R(t,t^\prime) G^>_{\vec{k}_1\downarrow}(t,t^\prime) G^>_{\vec{k}_2\sigma}(t,t^\prime) G^<_{\vec{k}^\prime\sigma}(t^\prime,t) 	+ \sum_{\vec{q}} V_{\vec{q}} V^\ast_{\vec{q}}\, D^R(t,t^\prime) G^<_{\vec{q}\uparrow}(t^\prime,t) \nonumber\\
	&+ \sum_{\vec{k}\vec{k}^\prime\sigma} V_{\vec{k}\vec{k}^\prime\sigma}(t) V^\ast_{\vec{k}\vec{k}^\prime\sigma}(t^\prime) S^R_{2\downarrow}(t,t^\prime) G^>_{\vec{k}\sigma}(t,t^\prime) G^<_{\vec{k}^\prime\sigma}(t^\prime,t)
	+ \sum_{\vec{k}\vec{q}\sigma} V_{\vec{k}\vec{q}\sigma}(t) V^\ast_{\vec{k}\vec{q}\sigma}(t^\prime) S^R_{2\sigma}(t,t^\prime) G^>_{\vec{k}\downarrow}(t,t^\prime) G^<_{\vec{q}\sigma}(t^\prime,t)~,\\
\label{SErS2}
	\Sigma^R_{2\sigma}(t,t^\prime) &= \sum_{\vec{k}} V^\ast_{\vec{k}\sigma}(t) V_{\vec{k}\sigma}(t^\prime) E^R(t,t^\prime) G^>_{\vec{k}\sigma}(t,t^\prime)
	+\sum_{\vec{k}} V_{\vec{k}-\sigma}(t) V^\ast_{\vec{k}-\sigma}(t^\prime) D^R(t,t^\prime) B^<_{-\sigma}(t^\prime,t) G^<_{\vec{k}-\sigma}(t^\prime,t)  \nonumber\\
	&+ \delta_{\downarrow \sigma} \sum_{\vec{k}\vec{k}^\prime\sigma^\prime} V^\ast_{\vec{k}\vec{k}^\prime\sigma^\prime}(t) V_{\vec{k}\vec{k}^\prime\sigma^\prime}(t^\prime) S^R_{1\downarrow}(t,t^\prime) G^<_{\vec{k}\sigma^\prime}(t^\prime,t) G^>_{\vec{k}^\prime\sigma^\prime}(t,t^\prime) 
	+ \sum_{\vec{k}\vec{q}} V^\ast_{\vec{k}\vec{q}\sigma}(t) V_{\vec{k}\vec{q}\sigma}(t^\prime) S^R_{1\downarrow}(t,t^\prime) G^<_{\vec{k}\downarrow}(t^\prime,t) G^>_{\vec{q}\sigma}(t,t^\prime)~,
\end{align}
\end{widetext}
where the physical origin of the various terms can again be identified by the type 
of the sum. 

\section{Auger levelwidths\label{AppWidth}}

In this appendix we indicate the main steps leading to the rate $\Gamma_{\rm AN}^<(t)$
in equation~\eqref{rateEq}. The other rates in this equation can be obtained 
similarly. 

As pointed out by Langreth and coworkers~\cite{LN91,SLN94a,SLN94b} the essential step for
obtaining~\eqref{rateEq} is to notice that the selfenergies 
are peaked around the time-diagonal. Hence, the time integrals 
in~\eqref{rateEqEa}--\eqref{rateEqDa} effectively set the time variables in the Green 
functions $E(t,t^\prime)$, $D(t,t^\prime)$, and $S_{n\sigma}(t,t^\prime)$--applying to the 
affinity and ionization levels of the projectile--to equal times (semiclassical approximation). 
Under the time integral of~\eqref{rateEqEa}, for instance, the function $iE^<(t,\bar{t})$ 
can be replaced by $n_+(t)$ while the function $iE^R(t,\bar{t})$ reduces to unity for $t>\bar{t}$ 
and vanishes otherwise. As a result of the semiclassical approximation, the quantum kinetic 
equations~\eqref{rateEqEa}--\eqref{rateEqDa} reduce to~\eqref{rateEq} with rates however not yet 
in a form numerically tractable.

For a numerical treatment of~\eqref{rateEq} the rates have to be simplified. Taking Auger
neutralization as an example we now explain the main steps of the simplification. The Auger
rate initially appearing in~\eqref{rateEq} follows from the selfenergy~\eqref{analyticContinuation}. 
It reads
\begin{align}
	\label{appRate}
	\Gamma^<_{\rm AN} (t) &= 2\,\mathrm{Re} \int_{-\infty}^{t} d\bar{t}\,\sum_{\vec{k}_1\vec{k}_2\vec{k}'\sigma}
 V_{\vec{k}_1\vec{k}_2\vec{k}'\sigma}^\ast(t) V_{\vec{k}_1\vec{k}_2\vec{k}^\prime\sigma}(\bar{t}) \nonumber\\ 
	&\times f^<(\varepsilon_{\vec{k}_1\downarrow}) f^<(\varepsilon_{\vec{k}_2\sigma}) 
        f^>(\varepsilon_{\vec{k}'\sigma}) \nonumber\\
	&\times\exp\bigg(i\int_{\bar{t}}^{t}d\tau\,\big(\varepsilon_{\vec{k}_1\downarrow}+
\varepsilon_{\vec{k}_2\sigma}-\varepsilon_{\vec{k}'\sigma}-\varepsilon^0_{1s\downarrow}(\tau) \big)\bigg)\,.
\end{align}
Adding an integration over $\varepsilon$ by inserting the delta-function  
$\delta (\varepsilon_{\vec{k}_1\downarrow}+\varepsilon_{\vec{k}_2\sigma}-\varepsilon_{\vec{k}'\sigma}-\varepsilon )$
we rewrite this expression as 
\begin{align}
\Gamma^<_{\rm AN}(t) &= 2\,\mathrm{Re} \int_{-\infty}^{t} d\bar{t}\,\int\dfrac{d\varepsilon}{2\pi}\,
\Gamma^{\rm AN}_\varepsilon(t,t^\prime) \nonumber\\
&\times \exp\bigg(i\int_{\bar{t}}^{t}d\tau\,\big(\varepsilon-\varepsilon^0_{1s\downarrow}(\tau) \big)\bigg)
\label{appRateNew}
\end{align}
with  
\begin{align}
\label{appKernel}
	\Gamma^{\rm AN}_\varepsilon(t,t') =  2\pi\sum_{\vec{k}_1\vec{k}_2\vec{k}'\sigma} &V_{\vec{k}_1\vec{k}_2\vec{k}'\sigma}^\ast(t) V_{\vec{k}_1\vec{k}_2\vec{k}^\prime\sigma}(\bar{t}) \nonumber\\
	\times\delta (\varepsilon_{\vec{k}_1\downarrow}+\varepsilon_{\vec{k}_2\sigma}-\varepsilon_{\vec{k}'\sigma}-\varepsilon )& f^<(\varepsilon_{\vec{k}_1\downarrow}) f^<(\varepsilon_{\vec{k}_2\sigma}) f^>(\varepsilon_{\vec{k}'\sigma})\,,
\end{align}
a function which can be simplified by noticing that the time and momentum dependencies of the Auger matrix 
elements $V_{\vec{k}_1\vec{k}_2\vec{k}^\prime\sigma}(t)$ and $V_{\vec{k}_1\vec{k}_2\vec{k}^\prime\sigma}(\bar{t})$ 
approximately factorize. As for single-electron transfer processes~\cite{LN91,SLN94a,SLN94b} we can thus 
approximately write
\begin{align}
\Gamma^{\rm AN}_\varepsilon(t,t') \approx \sqrt{\Gamma^{\rm AN}_\varepsilon (t) 
\Gamma^{\rm AN}_\varepsilon (t')}\,,
\label{GammaAuxNew}
\end{align}
where
\begin{align}
\Gamma^{\rm AN}_\varepsilon(t) &=  2\pi\sum_{\vec{k}_1\vec{k}_2\vec{k}'\sigma} 
\vert V_{\vec{k}_1\vec{k}_2\vec{k}'\sigma}(t) \vert^2 \nonumber\\
&\times\delta (\varepsilon_{\vec{k}_1\downarrow}+\varepsilon_{\vec{k}_2\sigma}-
\varepsilon_{\vec{k}'\sigma}-\varepsilon)\nonumber\\
&\times f^<(\varepsilon_{\vec{k}_1\downarrow}) f^<(\varepsilon_{\vec{k}_2\sigma}) 
f^>(\varepsilon_{\vec{k}'\sigma})\,
\label{GammaAux}
\end{align}
is essentially the levelwidth introduced in~\eqref{levelwidthAN} except that the 
energy $\varepsilon$ is not yet pinned to $\varepsilon^0_{1s\downarrow}$. This 
is accomplished by the time integration. Indeed, inserting~\eqref{GammaAuxNew} into~\eqref{appRateNew} 
and applying again a saddle-point approximation we obtain 
\begin{align}
\Gamma^<_{\rm AN} (t) &= 2\,\mathrm{Re} \int_{-\infty}^{t} d\bar{t}\,\int\dfrac{d\varepsilon}{2\pi}\, 
\sqrt{\Gamma^{\rm AN}_\varepsilon (t) \Gamma^{\rm AN}_\varepsilon (\bar{t})} \nonumber\\ 
&\times \exp\bigg(i\int_{\bar{t}}^{t}d\tau\,\big(\varepsilon-\varepsilon^0_{1s\downarrow}(\tau) \big)\bigg) 
\nonumber\\
&\simeq 2\,\mathrm{Re} \int_{-\infty}^{t} d\bar{t}\,
\sqrt{\Gamma^{\rm AN}_{\varepsilon^0_{1s\downarrow}(t)} (t) \Gamma^{\rm AN}_{\varepsilon^0_{1s\downarrow}(\bar{t})} (\bar{t})}\nonumber\\
&\times \int\dfrac{d\varepsilon}{2\pi}\, \exp\bigg(i\int_{\bar{t}}^{t}d\tau\,\big(\varepsilon-\varepsilon^0_{1s\downarrow}(\tau) \big)\bigg) \nonumber\\
&= 2\,\mathrm{Re} \int_{-\infty}^{t} d\bar{t}\,\sqrt{\Gamma^{\rm AN}_{\varepsilon^0_{1s\downarrow}(t)} (t) \Gamma^{\rm AN}_{\varepsilon^0_{1s\downarrow}(\bar{t})} (\bar{t})}\,\delta (t-\bar{t})\,,
\label{appLastTerm}
\end{align}
where the last line, when the time integral is carried out, yields 
$\Gamma^<_{\rm AN} (t) = \Gamma_{\rm AN}(t)$. The main gain numerically is that for 
$\Gamma_{\rm AN}(t)$ given by~\eqref{approxAN} it is only necessary to calculate the 
squared modulus of the Auger matrix element at the time appearing also in 
the rate equation~\eqref{rateEq} whereas in~\eqref{appRate} the matrix element
has to be determined also for all times past the actual time.

For single-electron transfer processes Langreth and coworkers~\cite{LN91,SLN94a,SLN94b} investigated 
in great detail the range of validity of the simplified rates. It depends on a number of conditions 
which are almost never rigorously satisfied. The original rates dropping out from 
Eqs.~\eqref{rateEqEa}--\eqref{rateEqDa} are however numerically too expensive to handle. 
From a practical point of view, the simplification described in this appendix seems to be 
unavoidable for producing numerical data. It has to be applied to all the rates of 
Eq.~\eqref{rateEq}. 

\bibliography{./ref.bib}
\bibliographystyle{apsrev}

\end{document}